\documentclass{article}
\usepackage[utf8]{inputenc}

\title{The Gossip Paradox: why do bacteria share genes?}
\author{Alastair Jamieson-Lane, Bernd Blasius}
\date{}

\usepackage{graphicx}
\usepackage{mathtools}
\usepackage{amssymb}
\usepackage{amsmath}
\usepackage{csquotes}
\usepackage{enumerate}
\usepackage{todonotes}
\newcommand{\PlasPlus}{\oplus} 
\newcommand{\PlasMinus}{\ominus}

\begin{document}

\maketitle

\abstract{Bacteria, in contrast to eukaryotic cells contain two types of genes: chromosomal genes that are fixed to the cell, and plasmids that are mobile genes, easily shared to other cells. The sharing of plasmid genes between individual bacteria and between bacterial lineages has contributed vastly to bacterial evolution, allowing specialized traits to `jump ship' between one lineage or species and the next. The benefits of this generosity from the point of view of both recipient and plasmid are generally understood, but come at the expense of chromosomal genes in the donor cell, which share potentially advantageous genes with their competition while receiving no benefit. Using both continuous models and agent based simulations, we demonstrate that `secretive' genes which restrict horizontal gene transfer are favored over wide range of models and parameter values. Our findings lead to a peculiar paradox: given the obvious benefits of keeping secrets, why do bacteria share information so freely?}

\section{Introduction}
Evolution, that blind process of inheritance and selection, is both a process of constant refinement and optimization, and also a process of adaption; while in stable environments the species best able to make efficient use of available resources may drive others to extinction, in more variable environments it is those species most able to change and adapt which will flourish.

This tension between efficiency and adaptability can be seen most clearly in bacterial genomes, where `core' genes are stored on bacterial chromosomes, while `accessory' genes are commonly found on mobile gene elements, such as plasmids: small stable loops of DNA, independent from the host chromosome.

Core genes code for critical metabolic pathways, cell division and motility, while accessory genes code for more esoteric capabilities- for example 
resistance to heavy metals \cite{summers_mercury_1972}, uncommon metabolic pathways\cite{shao_dna_2009},  virulence factors \cite{johnson_pathogenomics_2009}, or antibiotic resistance \cite{bennett_plasmid_2008}. Genes stored on mobile gene elements can be lost, rearranged, and most importantly for our purposes, passed from one bacteria to another via the process of Horizontal Gene Transfer (HGT).

Research into HGT over the past decades has fundamentally altered our understanding of bacterial evolution. Rather than a gradual accumulation of mutations within a single continuous clonal lineage, HGT allows entire gene modules to be transferred from one genome to another, fulfilling much the same role as sexual recombination. Bacterial genomes are less a tapestry which must be altered one thread at a time, but instead a patchwork quilt, with some estimates \cite{dagan_modular_2008} suggesting that upwards of $80\%$ of bacterial genes have undergone HGT at some point in their history. While the historical impact of HGT is enough to earn great scientific and philosophical interest, the role of HGT in the spread of antibiotic resistance and bacterial virulence factors also fuels significant medical and practical concerns in our understanding of HGT \cite{yang_conjugative_2019,chen_evolution_2019,peter_tracking_2020}.

\begin{figure}[p]
    \centering
    \centerline{ 
\includegraphics[width=1.05\textwidth]{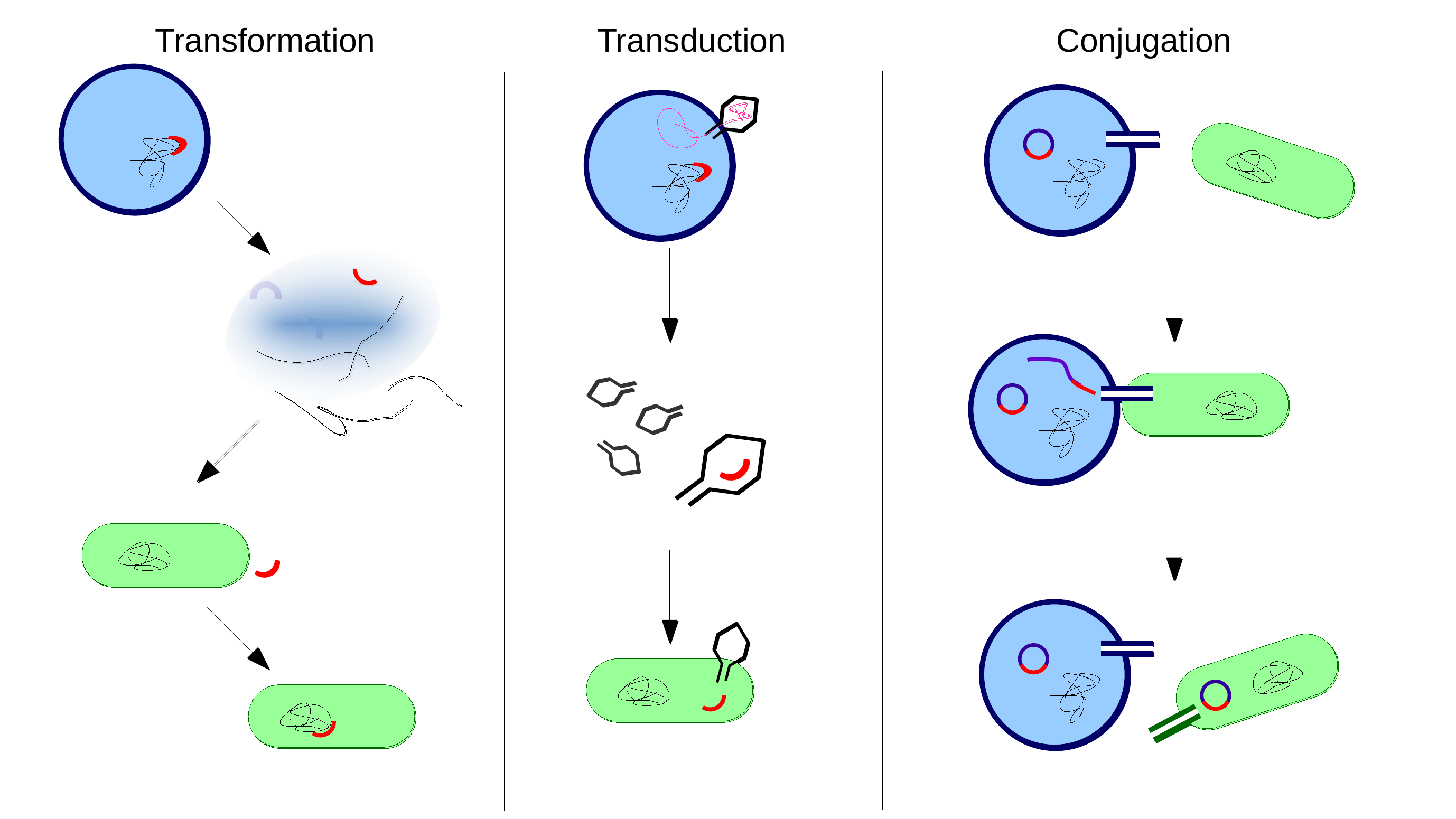}
}
     \caption{Modes of Horizontal Gene Transfer. Left: HGT via transformation. After a cell dies, the lysing cell spills fragments of genes into the environment. Later, some other cell collects one of these fragments, incorporating the gene fragment into its genome. Middle: HGT via transduction. A phage infects a cell, hijacking the cellular machinery in order to produce further phages. In the process, one of the new phages is accidentally constructed containing a piece of the original cell's DNA, rather than the phage's. Later, this defective phage inserts the DNA fragment into a new cell. Right: HGT via conjugation. A cell contains DNA on a plasmid. This plasmid codes for bacterial conjugation, leading to the construction of a `pilus'. The donor cell binds to another bacteria, duplicates the plasmid DNA, and sends it through the pilus to its new host. Afterwards, the two cells go their separate ways.}
    \label{fig:HGTcartoon}
\end{figure}

HGT can take one of several pathways (see Fig. \ref{fig:HGTcartoon}). It can take place via {\bf transformation}: one bacterial cell taking up genetic material from its environment \cite{johnson_integrative_2015}. The process of {\bf transduction} takes place when HGT takes place via an invading viruses. By chance, a virus may accidentally package host DNA inside a viral capsid before lysing a cell, eventually leading to DNA from one host being inserted into the next. Most strikingly of all, accessory genes can be shared via bacterial {\bf conjugation}, in which one cell bridges the gap to another, duplicates its plasmids, and funnels these copies across. For the interested reader seeking further detail we recommend Eberhard's classical review paper \cite{eberhard_evolution_1990}, and the more recent overview of Thomas and Nielson \cite{thomas_mechanisms_2005}.

Conjugation is extensively regulated, both from the point of view of the donor and recipient. On the side of the donor cell, plasmid bound quorum sensing genes determine the appropriate density of donor cells for a particular environmnet \cite{koraimann_social_2014} and intricate restriction-modification mechanisms (what might be thought of as the `bacterial immune system') act to suppress foreign gene elements \cite{oliveira_regulation_2016}, though under what conditions a such foreign elements are accepted and when they are repressed remains unclear. On the side of the reciepient cell, individual plasmids exclude incoming plasmids from the same `incompatibility group', as these are likely to disrupt plasmid replication and regulatory processes \cite{garcillan-barcia_why_2008}. Conjugation rate depends both on the plasmid in question, but also on the bacterial background these plasmids inhabit \cite{hall_sourcesink_2016}. 

Bacterial conjugation is, for many reasons, somewhat odd. In particular, it involves one bacteria supplying another bacteria with (presumably) beneficial genetic material, often at reasonable expense to itself \cite{dahlberg_amelioration_2003}. 
From the point of view of the plasmid genes being shared, this expense makes sense- the time and resources involved create more copies of the plasmid and allow it to ride over selective sweeps \cite{bergstrom_natural_2000} or spread to new species or environments \cite{bottery_adaptive_2017}. 
From the point of view of chromosonal genes in the recipient cell, the benefits are also clear; if the plasmid is beneficial the recieving bacteria and its descendants gain in fitness. Given the sometimes heavy burden imposed by newly arrived plasmid genes\cite{trautwein_native_2016,san_millan_fitness_2017,baltrus_exploring_2013}, the cost of accepting new plasmids can not be assumed to be trivial. Acceptance of an arriving plasmid can viewed as an act of `trust', but not one of `generosity' (at least, to the extent that such intentionalistic language can be applied to bacterial behavior). Though interesting in its own right, this question of trust will not be the focus of our present work.

While the benefits of HGT to the plasmid are clear, and the benefit to the receiving cell is at least somewhat understandable, from the perspective of the chromosomal genes in the donor cell, conjugation would appear to be strictly harmful.
Not only does conjugation take significant time and risk infection by certain types of phages \cite{colom_sex_2019}, in sharing adventageous DNA, a bacteria runs the risk of not only helping a rival in the present, but of generating a superior lineage and rendering its own decendants obsolete. In some cases, the recipient cell may even be of a completely different species altogeather \cite{norman_conjugative_2009}. 

While previous authors have studied the peculiar evolutionary forces resulting from HGT from the point of view of bacterial plasmids- asking such questions as `how are deleterous genes not lost?' and `why are beneficial genes not incorporated into chromosomal genomes more directly?' \cite{harrison_plasmid-mediated_2012}, in this work we instead consider HGT from the point of view of those chromosomal genes left behind. Our goal in what follows is to investigate and demonstrate the tension between the interests of chromosomal and plasmid bound genes, particularly with respect to `preferred' conjugation rate.

In section \ref{sec:OneSweep} we make use of a deterministic model to study the evolutionary dynamics in the simple case of one plasmid and two possible conjugation rates; we find that bacterial strains which suppress plasmid conjugation out compete their more generous competitors. We extend this model in section \ref{sec:ManySweep} in order to consider longer evolutionary time spans. In section \ref{sec:agentModel} we construct a more detailed agent based model, and through simulation demonstrate how genes that limit plasmid conjugation gain a long term evolutionary advantage over a wide parameter range, even when accounting for space. In section \ref{sec:comparePastModels} we build further upon this model, and discuss the similarities and differences between HGT and previous evolutionary `paradoxes'; namely the emergence of altruism and the evolution of sex. In both cases we find that classical resolutions to these past paradoxes fail to stabilize plasmid sharing.

\section{A Single Selective Sweep}
\label{sec:OneSweep}
Our goal throughout this article will be to explore the evolutionary pressures experienced by chromosomal genes and how they interact with more mobile plasmids. While the course of evolution is governed by many factors, in this section our goal is to construct the simplest possible deterministic model in order to study the interaction between mobile and `static' genes, and build our intuition, before moving on to more complex modelling approaches.

\begin{figure}[p]
    \centering
    \centerline{ 
\includegraphics[width=0.95\textwidth]{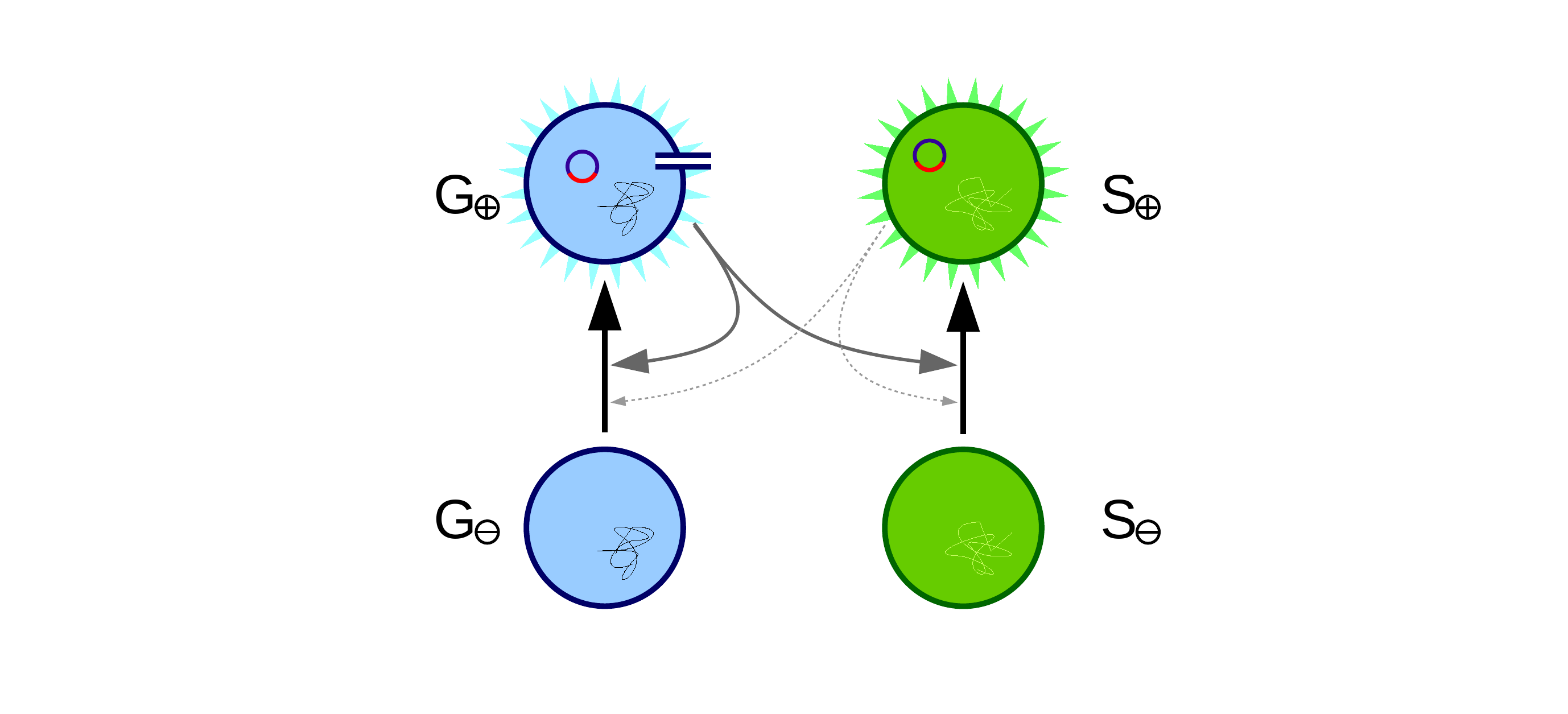}
}
     \caption{Our four `variants', $G_\PlasPlus,G_\PlasMinus,S_\PlasPlus$ and $S_\PlasMinus$. `Generous' variants are depicted in blue (left), `secretive' in green (right). Plasmids are actively spread by $G_\PlasPlus$ type bacteria, who posses both the plasmid and the chromosomal gene inclined to spread it. $S_\PlasPlus$ type bacteria share the plasmid at a much lower rate, potentially close to zero. All variants replicate according to their fitness, with plasmid carrying cells assumed to have $f_\PlasPlus>f_\PlasMinus$. The case of burdensome plasmids, or plasmids with varying fitness, we ignore for the time being.
     }
    \label{fig:SpeciesConsidered}

    \centering
    \centerline{ 
\includegraphics[width=0.85\textwidth]{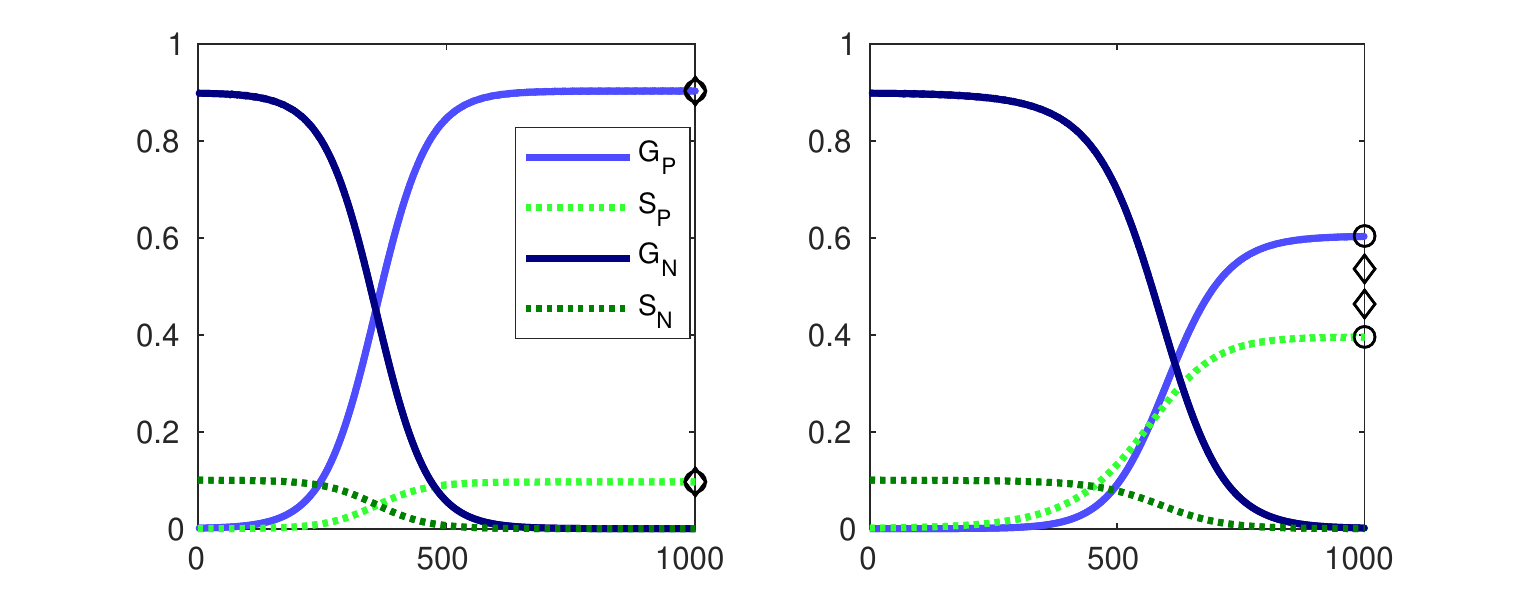}
}
     \caption{Solving equations \ref{eq:DetMain}, we observe an incoming plasmid arrive and become ubiquitous in the population (a selective sweep). As it does so, the balance between generous and secretive genes shifts. $c=0.01$, $\epsilon=0.01$, $f_\PlasPlus=1.01$, $f_\PlasMinus=1$ (Left) Initial plasmid positive population is $G_\PlasPlus=10^{-3}, S_\PlasPlus=0,G_\PlasMinus \approx 0.9, S_\PlasMinus=0.1$. (Right) Initial plasmid positive population is $G_\PlasPlus=0, S_\PlasPlus=10^{-3},G_\PlasMinus \approx 0.9, S_\PlasMinus=0.1$.
     In the right hang panel, black diamonds represent the predicted final values using the approximation eq.\ref{eq:Minf}, while circles represent predictions made by solving eq. \ref{eq:FindM_inf} using Newton's method. In the left hand panel, these values are indistinguishable and the symbols are placed one on top of the other. Overall, when starting with a generous mutant (left), the levels of generous individuals increases from $0.899$ to $0.9035$, a $0.5\%$ increase. When starting with a secretive mutant (right), the secretive population increases from $0.100$ to $0.3955$, a four-fold increase.
     }
    \label{fig:Sweep}
\end{figure}

Consider a population of bacteria, each of which is either positive ($\PlasPlus$) or negative ($\PlasMinus$) for some beneficial plasmid. In addition, each bacteria possess either a generous (G) or secretive (S) gene on their chromosome, which share plasmids via conjugation at rate $c$, or at some drastically reduced rate $\epsilon c$. Note that this genotype determines a bacteria's propensity to \emph{donate} plasmids to others; all bacteria are equally capable of receiving plasmids. Here $\epsilon c$ accounts for both imperfect conjugation suppression on behalf of our chromosome, and also `plasmid leakage' via the unregulated HGT processes of transduction or transformation.
Plasmid positive bacteria have some fitness $f_\PlasPlus$, while plasmid negative bacteria have fitness $f_\PlasMinus<f_\PlasPlus$. See figure \ref{fig:SpeciesConsidered} for a schematic of this set up.

Written as chemical equations we have:
\begin{align}
	G_{\pm} & \xrightarrow[f_{\pm}]{} 2 G_{\pm}, ~ &
	S_{\pm}  \xrightarrow[f_{\pm}]{}& 2 S_{\pm},\\
   G_\PlasPlus + S_\PlasMinus & \xrightarrow[c]{} G_\PlasPlus + S_\PlasPlus, ~ &  G_\PlasPlus + G_\PlasMinus  \xrightarrow[c]{}& 2 G_\PlasPlus \\
   S_\PlasPlus + G_\PlasMinus & \xrightarrow[\epsilon c]{} S_\PlasPlus + G_\PlasPlus,~ &  S_\PlasPlus + S_\PlasMinus  \xrightarrow[\epsilon c]{}& 2 S_\PlasPlus 
\end{align}

We assume a `death rate' such that total population is held constant; $G_\PlasPlus+G_\PlasMinus+S_\PlasPlus+S_\PlasMinus =1$  at all times.

Written as a series of differential equations we have:
\begin{equation}
\begin{array}{rcl}
\dot G_\PlasPlus &=& (f_\PlasPlus-\bar f)G_\PlasPlus + (c G_\PlasPlus + \epsilon c S_\PlasPlus) G_\PlasMinus,\\
\dot S_\PlasPlus &=& (f_\PlasPlus-\bar f)S_\PlasPlus + (c G_\PlasPlus + \epsilon c S_\PlasPlus) S_\PlasMinus,\\
\dot G_\PlasMinus &=& (f_\PlasMinus-\bar f)G_\PlasMinus - (c G_\PlasPlus + \epsilon c S_\PlasPlus) G_\PlasMinus,\\
\dot S_\PlasMinus &=& (f_\PlasMinus-\bar f)S_\PlasMinus - (c G_\PlasPlus + \epsilon c S_\PlasPlus) S_\PlasMinus.\\
\end{array}
\label{eq:DetMain}
\end{equation}

Here $\bar f = f_\PlasPlus(G_\PlasPlus+S_\PlasPlus)+f_\PlasMinus(G_\PlasMinus+S_\PlasMinus) $ refers to the average fitness of the current population. The first term in each derivative refers to changes in population due to population growth and death/displacement. The second term in each derivative gives the demographic change caused by plasmid conjugation. These equations, can be viewed as filling similar role to the standard evolutionary replicator equations \cite{nowak_evolutionary_2006,cressman_replicator_2014}, although HGT somewhat complicates this picture.

Suppose, we add a single beneficial plasmid to an otherwise plasmid negative population. This plasmid may find itself in either a generous or secretive bacteria. Due to its higher fitness, the plasmid is going to `sweep' through, becoming ubiquitous in our population (a so called ``selective sweep'').  An illustration of two such selective sweeps is given in figure \ref{fig:Sweep}. 
Given the interaction between plasmids and chromosomes, we wish to determine how the balance between generous and secretive chromosomes changes as the new plasmid reaches saturation in our population. Stated mathematically, we wish to determine $G_\PlasPlus(\infty)$ and $S_\PlasPlus(\infty)$, given our parameters and initial conditions.

For this task, it proves convenient to make use of a slight relabeling of variables.
Consider the population of mutants, $M$, which are always plasmid positive, and either all generous or all secretive; $M$ represents the original mutant bacteria and all its direct descendants. We also consider a ``Wild type'' population $W$, which can be plasmid positive $W_\PlasPlus$ or plasmid negative $W_\PlasMinus$. The wild type population consists of some mixture of generous and secretive chromosomal genes. Because $ \frac{d}{dt}{(G_\PlasMinus /S_\PlasMinus)}=0$ we know that the relative proportions of generous and secretive genes remains fixed for $W_\PlasMinus$. Because $\dot W_\PlasPlus$ grows proportional to $W_\PlasPlus$ in its first term, and $W_\PlasMinus$ in its second, and because $W_\PlasPlus$ initially has the same $G/S$ ratio as $W_\PlasMinus$, it must inevitably continue with this same ratio throughout the entire course of our ODE. Thus, the relative ratio of generous and secretive chromosomal genes are fixed in the wild type population.

Stated algebraically, we have:
\begin{equation}
\begin{array}{rcl}
G_\PlasPlus &=& \rho W_\PlasPlus+ \chi M,\\
S_\PlasPlus &=& (1-\rho) W_\PlasPlus+ (1-\chi) M,\\
G_\PlasMinus &=& \rho W_\PlasMinus,\\
S_\PlasMinus &=& (1-\rho) W_\PlasMinus.\\
\end{array}
\end{equation}
Where $\rho= \frac{G_\PlasMinus(0)}{G_\PlasMinus(0)+S_\PlasMinus(0)}$, and $\chi=\frac{G_\PlasPlus(0)}{G_\PlasPlus(0)+S_\PlasPlus(0)}$.
This partitioning of the population is algebraic rather than physical, but allows us to simplify our equations:

\begin{equation}
\begin{array}{rcl}
\dot M & = &(f_\PlasPlus - \bar f) M,\\
\dot W_\PlasMinus & = &(f_\PlasMinus - \bar f) W_\PlasMinus - (c_M M + c_W W_\PlasPlus)W_\PlasMinus,\\
\dot W_\PlasPlus & = &(f_\PlasPlus - \bar f) W_\PlasPlus + (c_M M + c_W W_\PlasPlus)W_\PlasMinus.
\label{eq:MutantTakeOver}
\end{array}
\end{equation}
Here $c_M$ and $c_W$ represent the conjugation rates of the mutant and wild type populations, and are given by $c_M=\chi c_G + (1-\chi) c_S$ and $c_W=\rho c_G +(1-\rho)c_S$.

\begin{figure}[h]
    \centering
    \centerline{ 
\includegraphics[width=0.95\textwidth]{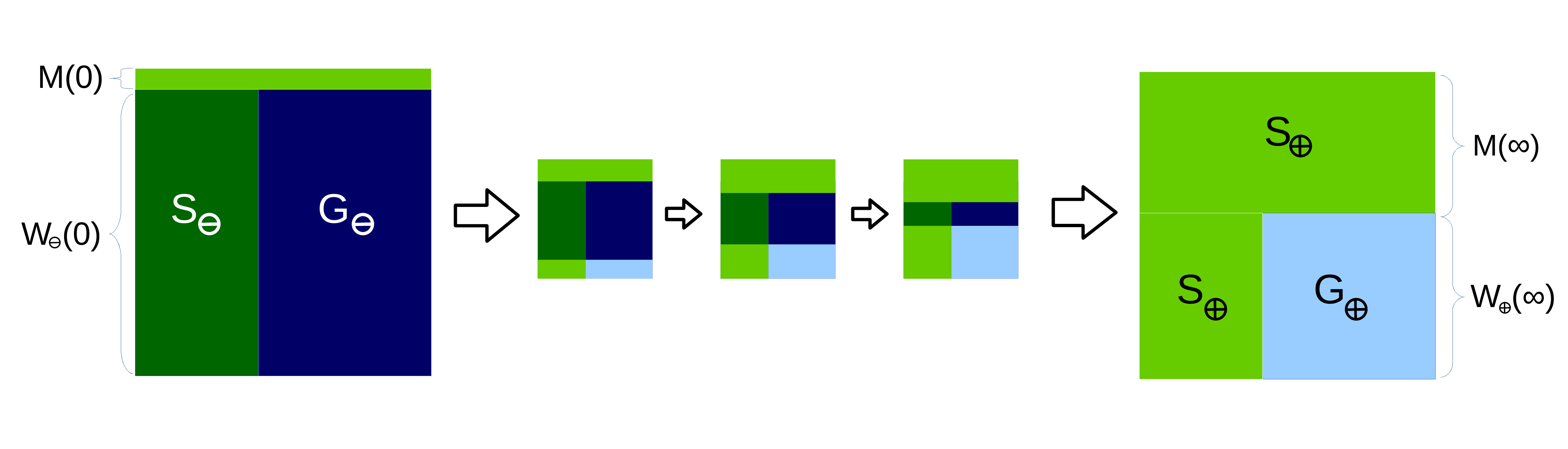}
}
     \caption{A graphical illustration of our change of variables. Presence/absence of beneficial plasmid is denoted by bright/dark colours. secretive/generous chromosomes are denoted by green and blue (respectively).  Over time descendants of our initial mutant displaces the wild type population,  while simultaneously, HGT via conjugation supplies the wild type population with the beneficial plasmid, leaving the relative abundance of the chromosomal genes in the wild type population unchanged. 
     }
    \label{fig:schematicChangeVariablesDiagram}
\end{figure}

As before, the total population is held constant, $M+W_\PlasMinus+W_\PlasPlus=1$, and $\bar f$ is the mean fitness of the current population, $\bar f= f_\PlasPlus M +f_\PlasPlus W_\PlasPlus + f_\PlasMinus W_\PlasMinus$.

Combining the definition of $\bar f$ with population conservation, we find that $(f_\PlasPlus - \bar f)= (f_\PlasPlus - f_\PlasMinus) W_\PlasMinus = \Delta f W_\PlasMinus$ and $(f_\PlasMinus - \bar f)= (f_\PlasMinus - f_\PlasPlus) (1-W_\PlasMinus) = -\Delta f (1-W_\PlasMinus)$.
Similar use of the total population equation gives
\begin{flalign}
\dot M  = & \Delta f W_\PlasMinus M,\label{eq:Mdot}\\
\dot W_\PlasMinus  =& -\Delta f (1-W_\PlasMinus) W_\PlasMinus - \left(c_M M + c_W (1-M-W_\PlasMinus) \right)W_\PlasMinus,\\
        =& -(\Delta f+c_W) (1-W_\PlasMinus) W_\PlasMinus - (c_M- c_W ) M W_\PlasMinus.
       \label{eq:Wdot}
\end{flalign}
Dividing the equation \ref{eq:Wdot} by \ref{eq:Mdot} eliminates time dependence:
\begin{align}
\frac{\dot W_\PlasMinus}{\dot M} & = -\frac{(\Delta f+c_W) (1-W_\PlasMinus)}{\Delta f M} - \frac{c_M- c_W }{\Delta f}.
\end{align}
The above can be rearranged to give a D'Alembert equation \cite{ince_ordinary_1956} of the form :
\begin{align}
W_\PlasMinus &=  \frac{( W_\PlasMinus'(M) + \beta) M}{\alpha} + 1,
\end{align}
where $\alpha = (\Delta f+c_W)/\Delta f$, $\beta=(c_M- c_W )/(\Delta f)$ and $W_\PlasMinus'$ is the derivative of $W_\PlasMinus$ with respect to $M$ (as opposed to $t$). This equation permits solutions of the form:
\begin{align}
W_\PlasMinus(M) = 1 + k M^\alpha + \frac{\beta M}{\alpha - 1},
\label{eq:W_of_M}
\end{align}
where $k$ is a constant of integration. Given the initial conditions are $M(0) = M_0 \ll 1$, $W_\PlasPlus=0$, $W_\PlasMinus=1-M_0$. This leads to:
\begin{align}
W_\PlasMinus(M_0)= 1-M_0 &= 1 + k M_0^\alpha + \frac{\beta M_0}{\alpha - 1},\\
\left(-1 -\frac{\beta}{\alpha - 1}\right) M_0&= k M_0^\alpha,\\
k &= -\left(1 + \frac{\beta}{\alpha - 1}\right) M_0^{1-\alpha}.
\label{eq:kVal}
\end{align}



The final mutant population is found by noting that $W_\PlasMinus \rightarrow 0$ as time proceeds, and hence
\begin{align}
W_\PlasMinus(M_\infty) =0 = 1 + k M_\infty^\alpha + \frac{\beta M_\infty}{\alpha - 1}.
\label{eq:FindM_inf}
\end{align}

If we assume that $M_0$ and $M_\infty$ are of the same order of magnitude, then we see that for $M_0 \ll 1$, the term $\frac{\beta M_\infty}{\alpha - 1}$ is negligible compared to the $+1$ term unless $\alpha \approx 1$. Hence
\begin{align}
M_\infty & \approx (-1/k)^{1/\alpha}, \\
  &\approx \left( 1 +\frac{\beta}{\alpha - 1}\right)^{-1/\alpha} M_0^{1-1/\alpha},\\
    &\approx \left( \frac{c_m}{c_w} \right)^{\frac{-\Delta f}{\Delta f +c_W}} M_0^{\frac{c_W}{\Delta f +c_W}}.
    \label{eq:Minf}
\end{align}

In cases where this formula suggests $M_\infty$ is large enough such that the approximation $M_\infty \ll 1$ fails, it is possible to instead use Newton's method to solve eq. \ref{eq:FindM_inf} numerically. Figure \ref{fig:Sweep} gives results using both methods.

For any beneficial plasmid, we can now determine the eventual population share that a mutant will have after one selective sweep, based on its initial fitness advantage, $\Delta f$, its propensity to conjugate and share plasmids, $c_m$, and the rate at which the (potentially mixed) wild type population shares plasmids $c_w$.

Evolution however, particularly of plasmids, does not consist of one selective sweep, but of many. This leads naturally to the question of `how do conjugation suppressing genes fare over many selective sweeps?'.


\section{Evolutionary Trajectory}
\label{sec:ManySweep}
Let us now consider how the balance of chromosomal genes evolves over many selective sweeps. 
Let $s_i$ refer to the fraction of the population containing $S$-type genes at the start of the $i^{th}$ selective sweep.
Each time a beneficial plasmid arrives (through mutation or immigration), it is acquired by an individual, `the mutant', which will be secretive with probability $s_i$ and generous with probability $1-s_i$. If the mutant is secretive, then by the end of the evolutionary sweep, secretive individuals will occupy $s_{i+1}=M_\infty + (1-M_\infty)s_i$ of the population. In contrast, if a generous individual is chosen the final population will have $s_{i+1}= (1-M_\infty)s_i$ - note that $M_\infty$ in each case will not be identical, as $M_\infty$ is calculated using (amongst other things) the conjugation rate of the mutant (see eq. \ref{eq:Minf}).

\begin{figure}[h]
    \centering
    \centerline{ 
\includegraphics[width=0.45\textwidth]{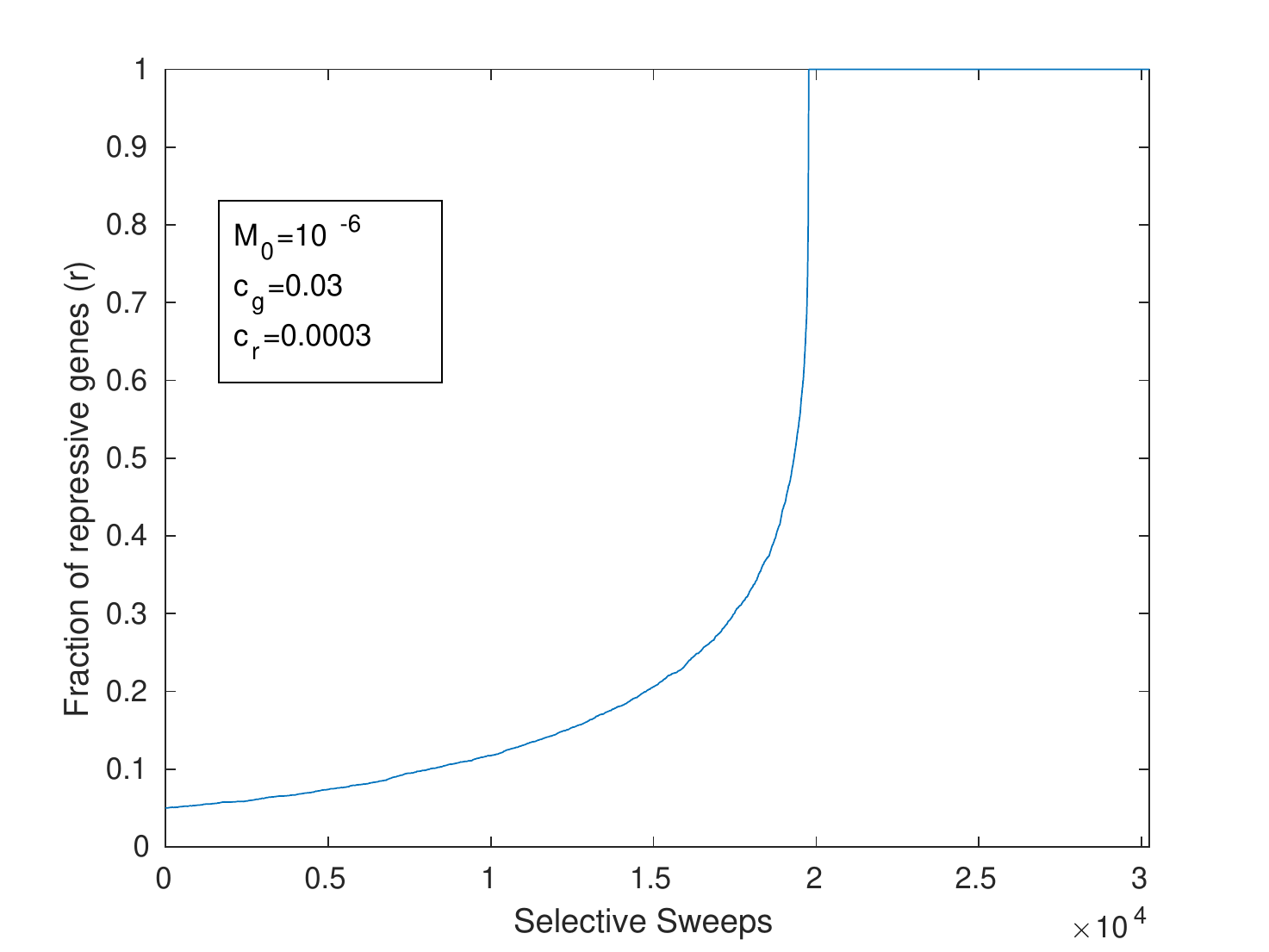}
\includegraphics[width=0.45\textwidth]{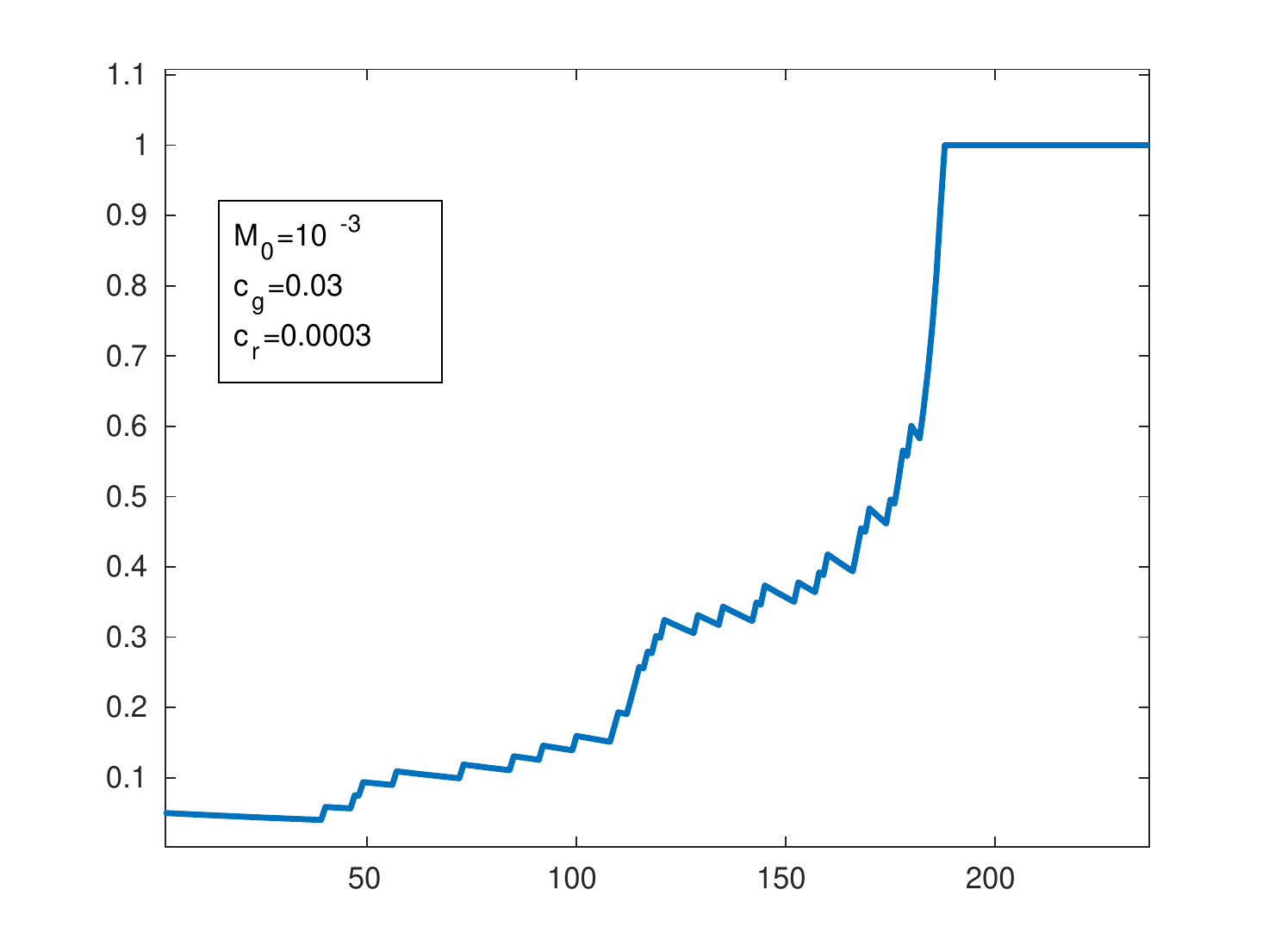}
}
     \caption{With each selective sweep the balance of the population changes. Over many sweeps, this leads to the fixation of lineages which are less inclined to share plasmids. However, for large population sizes (Left) this generally takes a very very long time, indicating that this effect is probably entirely overpowered compared to the many many other effects influencing evolution. (Right) we repeat the same, but with smaller population size/ larger $M(0)$, so as to make the individual jumps more visible.}
    \label{fig:SelectiveSweeps}
\end{figure}

Figure \ref{fig:SelectiveSweeps} shows lineages with low conjugation rate taking over a population; this take over is initially very slow (taking thousands of evolutionary sweeps), accelerating as secretive genes take over and overall conjugation rate is reduced. 

In a population with a generally high rate of conjugation ($c_w \approx \Delta f$), almost any non zero $c_m$ will quickly lead to plasmids spreading through the population, hence the advantage enjoyed by secretive lineages is minor. If conjugation rates in a population are already low secretive chromosomal genes enjoy a far greater advantage, reaching high population density after only a few selective sweeps. Regardless of the exact parameters however, given sufficient time, secrecy is favored. 

\section{An Agent based model}
\label{sec:agentModel}
The simplified evolutionary dynamics considered in the section \ref{sec:OneSweep} indicate that chromosomal genes which repress plasmid spread gain an evolutionary advantage (however slight) over long evolutionary time frames. While useful for the sake of analytic tractability and building intuition, this simplistic model ignores many important features, several of which may contribute to stabilizing plasmid sharing.
In order to test the robustness of our previous findings, let us now consider a more detailed ``agent based'' model. We will start by constructing a `baseline' model, and then will consider a number of extensions.

Consider a population of bacteria on an $N$ by $N$ grid with periodic boundary conditions. Each grid cell contains a single bacteria which is either generous ($G$), secretive ($S$) or dead ($X$). In nature, plasmids exist in a number of `incompatibility groups' \cite{thomas_plasmid_2021,cullum_rate_1979,scott_regulation_1984} - plasmids belonging to the same incompatibility group make use of the same regulatory proteins, and thus interfere with one anothers reproduction, driving one another to extinction whenever they exist within the same cell. Plasmids from different incompatibility groups co-exists with no such interference. 
In our model, we assume that each living bacteria possess one of $k$ possible plasmids from each of $q$ incompatibility groups. If there are (for example) $q=3$ incompatibility groups, the state of a single grid cell is given by a vector ${\bf v}$ of the form ${\bf v}= \{G,2,5,3\}$. 

Individual cells die at some death rate $d({\bf v},t)$, and reproduce at some reproduction rate $r({\bf v},t)$, both of which depend on the plasmids contained within the cell, and the current time. Dying cells will transition to the state ${\bf v}= \{X,-,-,-\}$. Reproducing cells will select a direction, and duplicate their state into the adjacent cell in the chosen direction, assuming the target cell is empty. If the selected cell is occupied, replication fails; this gives a crude representation of competition.

Plasmid mutation within cells occurs at some low constant rate $m=10^{-5}$; cells undergoing mutation select a plasmid group randomly, from $1$ to $q$, and set its value to some random value from $1$ to $k$. This mutation process is included so as to prevent plasmid extinction and ensure a minimal level of variance in the population.

Finally, cells conjugate at a rate $c({\bf v})$; for the purposes of our current exploration, conjugation rate depends only on chromosonal genes, with default rates $c(G)=0.1$, $c(S)=10^{-3}$ and $c(X)=0$. A conjugating cell selects a neighboring gridcell uniformly at random, and replaces one of the neighbors plasmids with its own value. If the selected gridcell is empty, conjugation fails. While it is well known that many plasmids possess surface exclusion mechanisms in order to prevent invasion by incompatible plasmids \cite{garcillan-barcia_why_2008}, these exclusion mechanisms are only partially effective \cite{gago-cordoba_surface_2019}. For the purposes of our current investigation, we assume that $c({\bf v})$ is the conjugation rate after taking pre-existing plasmid exclusion mechanisms into account.

All events are assumed to be exponentially distributed, and we simulate the entire system using Gillespie's Algorithm. Simulation code is avaliable via Github \cite{}.

\begin{figure}[h]
    \centering
\includegraphics[width=1.05\textwidth]{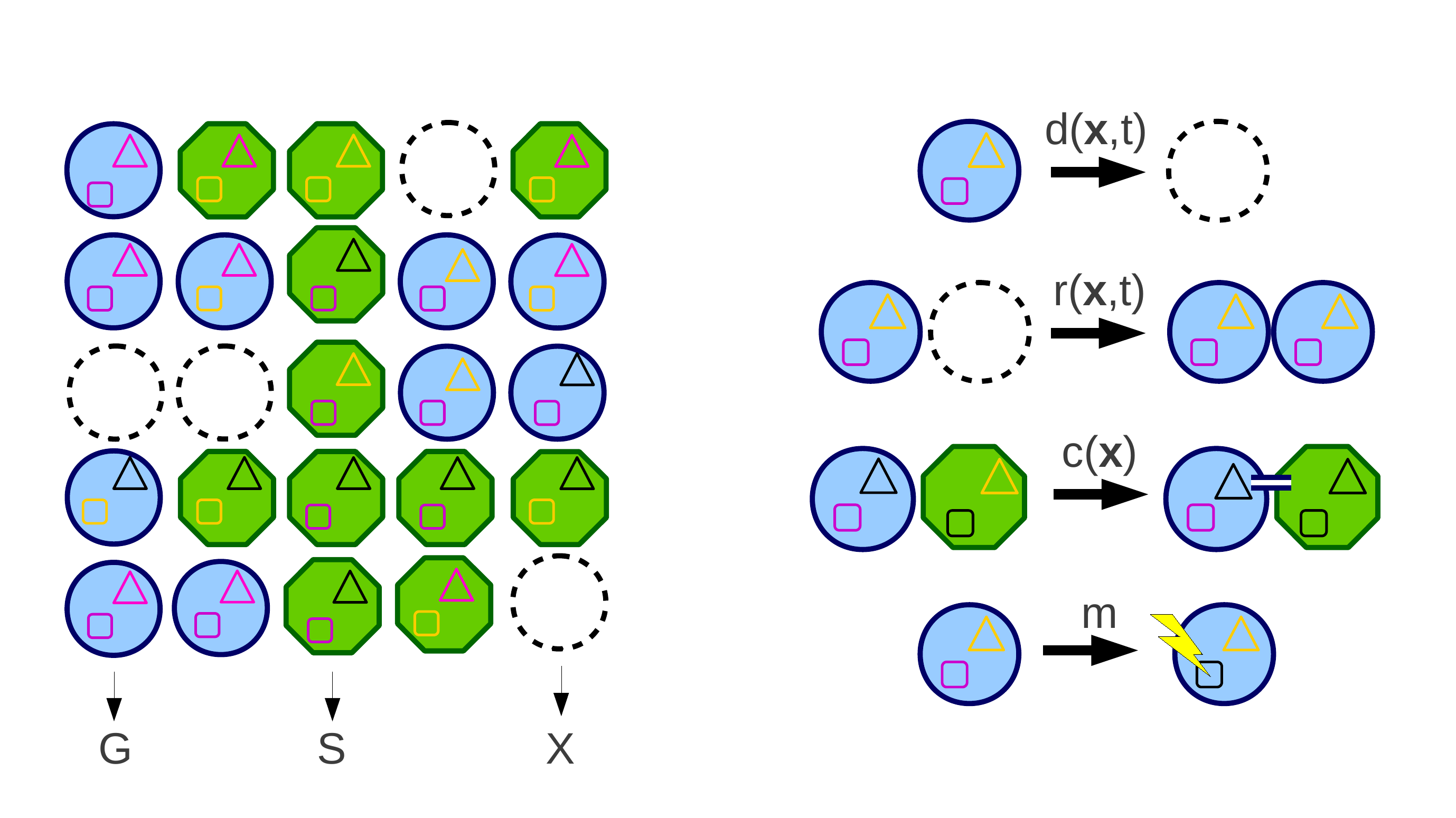}
\caption{(Left) An N by N grid ($N=5$) with periodic boundary conditions containing `generous' type bacteria (blue cricles), secretive bacteria (green octagons) and empty space. In this example, each bacteria contains a plasmid belonging to $k=2$ incompatibility groups (one group represented by triangles, the other by squares). Each plasmid comes in one of $q=3$ different variants (variants distinguished by colour). (Right) The system evolves according to four mechanisms: death, birth, conjugation and mutation.}
\label{fig:IntroGrid}
\end{figure}

A schematic illustration of this model is given in figure \ref{fig:IntroGrid}.

\subsection{Base case}
The agent based model described above is rather detailed. For the sake of concreteness, let us begin our investigation by considering the case where each bacteria contains only a single plasmid ($q=1$), and this plasmid comes in one of $k=4$ varieties. Assume $N=64$. Each gridcell initially has a $54\%$ chance of containing generous bacteria, $6\%$ secretive bacteria and 40\% empty space. The plasmid type for each bacteria in the initial population is selected uniformly at random. 

At time $t=0$, the simulation selects $d({\bf v},t)$ and $r({\bf v},t)$ uniformly at random in the ranges $(0.35-0.42)$ and $(1-1.3)$ respectively, for each plasmid state. At time $t=\tau=50$, and every  $\tau$ minutes thereafter, new death and replication rates are selected for each plasmid state. We refer to each such time window as an `epoch' and refer to $\tau$ as the `epoch time'. This approach gives a rudimentary approximation of the manner in which plasmid fitness changes with variable environmental conditions.

\begin{figure}[h]
    \centering
\includegraphics[width=0.95\textwidth]{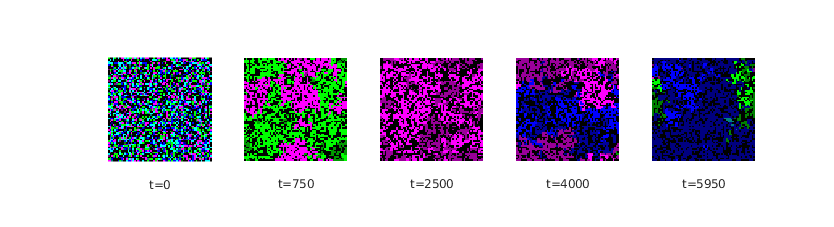}
\includegraphics[width=0.95\textwidth]{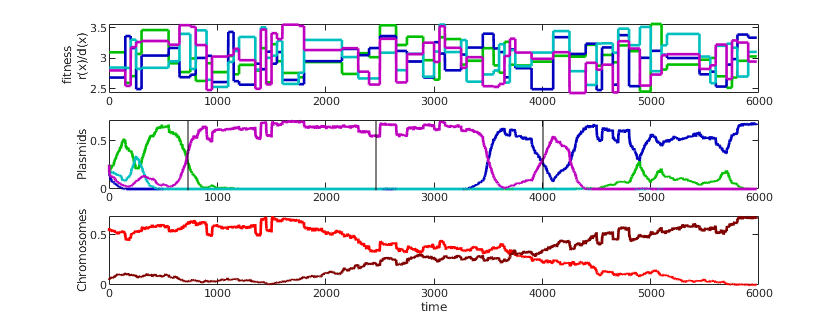}
\caption{(Above), illustrations of the full system state at a variety of times. Pixel colours give the identity of the corresponding plasmid, with bright colored nodes being generous, and darker nodes secretive. (Graphs, top) Every $\tau$ time steps, new $r({\bf v})$ and $d({\bf v})$ values are selected, leading to different `fitness' values $r({\bf v})/d({\bf v})$ for each of the four plasmid types. (Graphs, middle) The abundance of each plasmid strain rises and falls throughout the simulation. (Graphs, bottom) Here we graph the abundance of generous and secretive chromosonal genes (bright and dark, respectively). Over time, secrecy is favored.}
\label{fig:OnePlasmidGrid}
\end{figure}

Illustrations of the systems state throughout one simulation run are given in figure \ref{fig:OnePlasmidGrid}. 
Simulation is allowed to run until either $t=20,000$, or $G$ or $S$ is driven to extinction.

We run $100$ simulations using the parameter values described. 
In 47 out of 100 simulations, despite their low initial population, secretive chromosomes reach fixation in the population, driving generous chromosomes to extinction. This rate is nearly five times higher than we would expect for a neutral mutation, which would reach fixation with $10\%$ probability (as they represent $10\%$ of the initial non-empty population). Generous chromosones reach fixation in 53 of 100 simulations; all simulations complete before $t=20,000$. 
In cases where secretive chromosomes reach fixation in the population, the average fixation time is $t_{fix}= 3024$ (60.5 epochs). The average time until secretive chromosomes reach extinction time is $t_{ext}= 1203$ (24 epochs). 
What these results indicate is that simple `network reciprocity' (discussed later) is insufficient to stabilize conjugation, at least, not for the simple 2d lattice considered here.

\subsection{Exploring parameter space, and environmental heterogeneity}

With the base case now covered, let us now consider a range of different simulation conditions. Our aim in what follows is to determine the robustness of our previous results, and determine which aspects of the system (if any) affect fixation probability of secretive genes. In all simulations, we assume $N=64$, and initially gridcells are empty with probability $40\%$ probability, generous with $54\%$ probability and secretive with $6\%$ probability. Table \ref{tab:sims} summarizes the results of this section. 

We start with two `sanity tests' on our simulation. First, We consider a `control condition', in which $c(S)=c(G)=0.1$. In this neutral case we expect a fixation probability for $S$ precisely equal to the initial population fraction, $\frac{6}{60}$. In practice, we observe $12\%$ fixation, $81\%$ extinction, and $7\%$ timeout (Scenario \# B, table \ref{tab:sims}). This is consistent with what we would expect by chance. 
Next, in order to test sensitivity to simulation geometry, we reuse the same parameters as the base case simulation, but allow bacteria to conjugate and replicate diagonally, such that each cell is adjacent to 8 others rather than 4 (scenario C). No substantial change in fixation probability is observed; this is reassuring, as it indicates that our results are not overly sensitive to arbitrary choices made in the construction of the simulation.

With these basic tests out of the way, let us now explore variations of the more biologically meaningful parameters.
We consider increasing/decreasing epoch times to $\tau=250$ and $\tau=5$ (corresponds to considering a more or less stable environment), decreasing/increasing conjugation rates to $c(S)=10^{-5}$ and $c(S)=10^{-2}$ (stricter and looser plasmid suppression, respectively), and setting the mutation rate a factor of $30$ higher or lower. In {\it all} cases, fixation probability for $S$-type genes remains high (scenarios DEFGHI). 

We also consider the case of either increasing death rate to $d({\bf v})=(0.75-0.90)$, dramatically reducing population density, or decreasing it to $d({\bf v})=(0.05-0.06)$, drastically increasing population density. In the former case, the fixation probability of $S$ type chromosones drops to $11\%$ -- close to what we would expect in the neutral case. This suggests that when populations are sparse enough to render plasmid conjugation rare and ineffective, plasmid suppressing genes have no evolutionary effect, either positive or negative (scenario J). In the low death rate/high density case, the most common result of simulations is timeout; no meaningful conclusions can be drawn in this parameter regime (scenario K). If we simplify our model by assuming full population density, and that all death events are immediately followed by a corresponding birth event (treating our system as a death-birth Moran process on a graph \cite{moran_random_1958,Lieberman_evolutionary_2005}) simulation speed can be significantly increased, and the timeout condition at $t=20,000$ can be removed. In this case fixation occurs in $46\%$ of cases, close to our base rate. When fixation does occur, the average time period is $t\approx 44,000$. Extinction occurs by $t \approx 17,400$ on average.
Taken together, these simulations indicate that the advantage of secretive type chromosomal genes is exceptionally robust to changes in parameter values. If we desire a system that will stabilize sharing of plasmids, simple changes to parameter values are insufficient.

\begin{figure}[h]
    \centering
\includegraphics[width=0.85\textwidth]{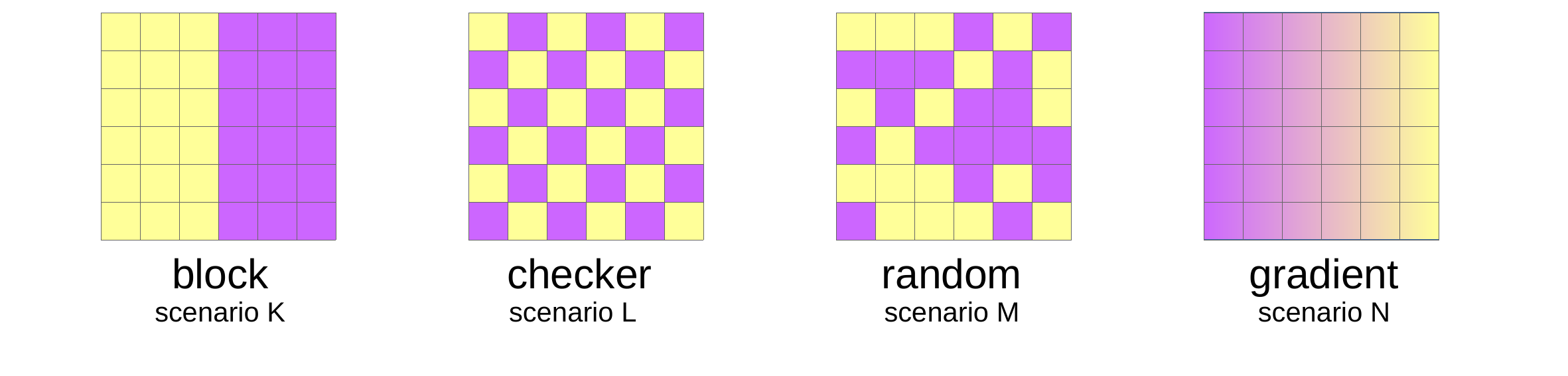}
\caption{Four possible environmental geometries; from left to right we have `block', `checkerboard',`random' and `gradient' type geometries. For each block, checkerboard and random conditions, cells experience one of two death and reproduction rates ($r_a({\bf v})$ or $r_b({\bf v})$) depending on the color of their gridcell. This allows us to simulate the behavior of cells around boundary regions (of various shapes). Block geometry assigns environment $A$ to all cells in the left half of the simulation and environment $B$ to the right half. Checkerboard assigns opposite environment conditions to adjacent cells. In the random condition each gridcell has a $50\%$ chance of having each environment type, each gridcell is determined independently. In the gradient condition, $r({\bf v})$ and $d({\bf v})$ are a linear combinations of the death and reproductive rates that would be experienced in `pure' environments.}
\label{fig:enviros}
\end{figure}

In natural settings, plasmids frequently contain resistance \cite{bennett_plasmid_2008} and metabolic \cite{shao_dna_2009} genes adapted for specific environments. For this reason we might expect plasmids to have particular evolutionary importance in and around transition regions or boundaries between different environments \cite{hermsen_rapidity_2012}.
In order to explore the effects of environmental variability, we extend our basic model to one in which each gridcell is assigned an `environment' parameter (either $A$ or $B$). Death and reproduction functions are replaced with environment specific functions $r_A({\bf v}), r_B({\bf v}), d_A({\bf v}), d_B({\bf v})$; with each epoch, the rate functions associated with either $A$ or $B$ (but not both) are changed.
In order to explore the effects of variable environment, we consider four different environmental geometries: block, checkerboard, random and gradient (see figure \ref{fig:enviros}). Death and replication rates in the gradient geometry are given as position dependent linear combinations of those found in $A$ and $B$ (hence $r({\bf v})=x r_A({\bf v})+(1-x) r_B({\bf v})$ for some position dependent $x$ in the range [0,1].
Regardless of the environmental condition considered, fixation is observed to occur in roughly $40-50\%$ of all cases (scenarios KLMN). Hence, it would appear that environmental heterogeneity is {\it also} insufficient to stabilize plasmid sharing.

\section{Relation to past paradoxes}
\label{sec:comparePastModels}
The paradox posed so far by the maintenance of conjugation is by no means the first paradox in the study of evolutionary dynamics, nor is it likely to be the last. Two paradoxes of the past, namely, the evolution of sex, and the evolution of altruism (also called the paradox of cooperation), bare striking resemblance to the current conundrum. Let us take a brief detour to examine each of these paradoxes, their resolutions, and the similarities and contrasts to the question currently under study.

\subsection{Can recombination preserve HGT?}
In the study of the evolution of sex, simple modelling suggests that asexual mutants would posses two major advantages over variants that reproduce sexually. Firstly, all members of the asexual species are able to reproduce, in contrast, for sexual species, only the female population can reproduce (the cost of males)\cite{maynard_smith_what_1971}. At the individual level, asexual individuals are able to pass on 100\% of their genes to each offspring, while sexual individuals pass on only 50\% of their genes (the cost of meiosis) \cite{williams_sex_2020, williams_introduction_1971}. Classically, this paradox is resolved by considering `the evolution of evolvability' \cite{institute_artificial_1989}. For asexual species, if beneficial mutants arise independently in two different lineages, one will inevitably drive the other to extinction, a process known as clonal interference. Clonal interference severely limits the speed of evolution, especially for large population size \cite{gerrish_fate_1998}. In contrast, sexual recombination causes beneficial genes to accumulate. Mating allows genes initially present in seperate lineages to come together in a single organism, both advantages can be retained, and the rate of evolution scales with population size \cite{otto_evolution_1997,colegrave_sex_2002}. 

Similar to sex, HGT allows genes to accumulate, and recombine. In contrast to sex, in which all genes have (approximately) equal chance of transfer and recombination, for bacterial conjugation, plasmid bound genes and other mobile gene elements gain the benefits of recombination, while core chromosomal genes do not. 
This leads to the question; are the benefits of recombination enough to stabilize plasmid sharing in a bacterial population?
In order to examine this we consider a collection of cells each containing $q=3$ plasmids, each from a different incompatibility group. For this simulation, we assume that individual conjugation events transport only a single plasmid, and thus, at the boundary between two clonal lineages, conjugation will quickly lead to a wide variety of different plasmid combinations, if cells are generous. This effect will be significantly reduced for secretive variants, leading to increased clonal interference.

In order to study the multi-plasmid case, we must first determine how $r({\bf v})$ is defined as a function of the three plasmid values $v_1$, $v_2$, $v_3$.
We consider three possible cases, in order of increasing complexity: in the first case (scenario Q), we assume that $r({\bf v})$ for each possible plasmid combination is selected in the range $(1-1.3)$ entirely independently, and that new $r({\bf v})$ are selected in each epoch. The fitness of the combination $[G,1,1,1]$ and that of $[G,1,1,2]$ are entirely independent of one another. One can think of this as being a `complex' genespace, the value of each plasmid varies based on the presence and absence of other plasmids. In the second case (scenario R), $r({\bf v})$ is formed as a linear combination $r({\bf v})= 1 +0.1 r_1(v_1)+0.1 r_2(v_2)+0.1 r_3(v_3)$. In this case $r_i(v_i)$ are selected in the range $(0,1)$. Each plasmid contributes to fecundity independently, and bacteria can succeed by `optimizing' for each plasmid incompatibility group independently. Finally, we consider the hybrid case, in which half of the fitness is determined via each of the previous two methods (scenario S).
In order to observe the benefits of recombination and the cost clonal interference, mutants must arrive frequently enough such that multiple mutant strains are competing at any one time. In order to achieve this, we assume $N=128$ and $m=10^{-4}$ for this batch of simulations.

In all three cases we observed significantly more fixation events than the control case (Scenario B), $45\%,47\%$ and $51\%$, respectively. Fixation and extinction times are comparable to the base case (Scenario A). Taken together these results indicate that recombination is also insufficient to maintain plasmid sharing (at least at the scale simulated here). The evolutionary pressure on the static chromosomal genes is too great. 

\subsection{Is HGT a public good?}
If the advantages of recombination are not enough to stabilize plasmid sharing, let us turn our attention to another evolutionary paradox, and its resolutions.
The second major paradox in the study of evolutionary dynamics is the emergence and maintenance of cooperation: slime molds cooperate in order to form stalks and bud \cite{ostrowski_enforcing_2019}, wolves hunt in packs \cite{cordoni_back_2019}, meerkats keep watch, and humans cooperate on vast and complex scales spanning the entire globe \cite{verma_revealing_2014} and beyond \cite{bell_interstellar_2015}. Cooperation is powerful, and yet at every turn, it is the organism that most benefits {\it its own} lineage that evolution will select. The lazy wolf, the cowardly meerkat, the spore that forces itself into the fruiting body rather than the stalk. Evolution is a game with winners and losers, not one that played for fun. Given the ever present advantage of selfish behavior, how then is cooperation maintained?- this is the question posed in the study of cooperation and public goods.

Numerous answers to this question have been proposed \cite{nowak_five_2006}, and at it would seem possible that several of the mechanisms that stabilize cooperation might also stabilize HGT. Much like cooperative and altruistic behavior, HGT involves one individual paying some `cost' for the benefit of another; how this benefit is distributed between the receiving cell and the plasmid itself is unknown, and likely to vary wildly between contexts. Unlike classical public goods interactions however, HGT involves the transfer not of resources, of {\it information}, a collection of genes that will potentially benefit the recipient not only in the immediate future, but for generations to come. When estimating the `value' of this gift, it is unclear what time horizon is appropriate.

With this differences and similarities in mind, let us now consider a variety of mechanisms that have been proposed for stabilizing cooperation, and how they might apply in the context of HGT. We note that there remains some debate in the literature as to the level of overlap between these mechanisms \cite{lehmann_group_2007,birch_kin_2015}. Here, we make no claims as to the distinctness or similarities between alternative evolutionary mechanisms, and instead simply err on the side of inclusivity whenever doing so appears physically appropriate in the bacterial context.

The most well known and well understood mechanism for stabiliztion of cooperation of evolutionary time scales is kin selection \cite{haldane_causes_1932}. 
Kin selection suggests that the benefits of cooperation are disproportionatly directed at ones relatives. Cooperative genes thus persist because they inevitably end up helping {\it other copies} of those same cooperative genes. This is in some sense similar to parental care, albeit on a wider scale. In the context of HGT, kin selection plays a strange role: those individuals most closely related to a plasmid donor are also those most likely to {\it already posses} a given plasmid, and hence gain no benefit. However, if a related individual {\it does not} posses a plasmid that the donor bacteria is currently benefiting from, they are also more likely than other bacteria to benefit, due to the plasmid and their chromosome being more closely adapted to one another \cite{dahlberg_amelioration_2003,stalder_emerging_2017}. The kin selection hypothesis is further hampered in the bacterial context, as there is little evidence that bacteria are able to track their kin (beyond spatial proximity), unlike vertebrate animals. Given that our previous simulations already include a spatial component, it would appear kin selection via spatial associativity is ineffective.

A second resolution that has been proposed to the paradox of cooperation, is reciprocity; that is to say, an individual who pays a cost today may well be on the recieving end of generosity tomorrow. Reciprocity comes in a variety of forms. Direct reciprocity involves two parties directly benefiting one another, such as plants supplying root fungi with sugars, in exchange for key nutrients\cite{kiers_reciprocal_2011,Hortal_role_2017}. Indirect reciprocity takes place when individuals provide benefit too, and gain benefit from a wider community; I am willing to help you because I trust that someone else will help me \cite{Nowak_evolution_2005, ohtsuki_how_2004}. Finally, `Network' reciprocity involves repeated interactions between neighboring individuals; these individuals are also likely to be related to the donor.
Neither direct, indirect nor network reciprocity would appear relevant in the context of bacterial conjugation. To the best of our current knowledge, bacteria are incapable of tracking the complex reputational networks required for indirect reciprocity, nor are they selective in who they donate to. Because export of plasmids depends on a variety of protein complexes within the donating cell \cite{sun_pull_2018}, plasmid donation is an explicitly `single directional' process, and hence direct reciprocity would also appear unlikely. Network reciprocity is already baked into the agent based model we are using, and has proven ineffective at preventing the invasion of secretive chromosomal genes.

One final resolution to the paradox of cooperation is the hypothesis of `group selection'. Under this model, individuals are separated into $M$ groups, and while selfish genes have the advantage on an individual level within each group, groups with a greater proportion of cooperative individuals are more likely to spread and divide than groups containing more selfish individuals. Unlike reciprocity and kin selection, this hypothesis is particularly suited to the microscopic world of bacteria. In order to test group selection in the context of HGT, we consider an alternative simulation, in which each column of the grid is considered one `group', and individuals may reproduce and conjugate to any cell within their column. Group selection can be implemented in a variety of different ways; for a classical review, see Wade 1978 \cite{wade_critical_1978}). For the sake of this initial exploration, we mimic Traulsen and Nowak's minimal model\cite{traulsen_evolution_2006}; cells duplicate within a group according to their fitness, expanding into the free space available, until eventually the group reaches size $N$. At this stage, with each duplication event, the group will either split with probability $w$, or a randomly selected group member will perish with probability $1-w$. When a group splits, one of the other groups is selected at random and is eradicated, and each member of the splitting group either migrates to the new space, or stays put (50\% probability of each outcome). Unlike previous simulation scenarios, death is not explicitly modeled, and is treated as a downstream consequence of either group splitting or individual replication. Conjugation is allowed to occur within groups, but not between groups; for the sake of conjugation, all groups are considered well mixed. 

In terms of simulation, we consider two separate scenarios. In both cases, we assume $w=10^{-4}$ (rare group splitting). For scenario U, we assume initial conditions $54\%$ generous, $6\%$ secretive (as previously). This results in generous chromosomes overtaking the population $91\%$ of the time, and secretive chromosomes reaching fixation $9\%$ of the time, a result in line with neutral selection.
In order to investigate this further, we also consider the same simulation with `balanced' initial conditions with initial populations of both generous and secretive chromosomes at $30\%$ (scenario V). Once again  we find results consistent with neutral selection. When plasmids are unable to transmit between groups, the advantage of hoarding them is neutralized, and fixation probability reflects initial population fractions almost perfectly.

As is clear from these experiments, the selective advantage of conjugation suppressing genes is robust across a wide range of parameter regimes, including regimes that have previously been found to stabilize cooperation or sexual reproduction. Fixation probability is only reduced back to the $10\%$ probability expected in for neutral evolution in circumstances where plasmid transmission is in some sense disabled (group isolation, sparse populations). Generous chromosomes are not found to have an evolutionary advantage for {\it any} of the scenarios considered.

We also observe that once fixation of secretive chromosomes is likely, fixation probability itself is relatively unaffected by parameter values. The general uniformity in fixation probability across scenarios would appear to indicate that extinction probability is governed almost entirely by local processes, and that larger scale and longer time frame effects (environment, mutation, epoch time), have little to no effect on the local dynamics. Given that extinction (when it occurs) happens 2-3 times faster than fixation, this may be indicative that fixation dynamics of secretive chromosomes are governed primarily by the probability of `early extinction', and that and once early extinction is avoided, fixation occurs with high probability.
The one exception to this `uniform fixation probability' is scenario F, in which we consider $c(S)=10^{-2}$, that is to say significantly weakened HGT suppression reduces the evolutionary advantage of. As might be expected in this case fixation of secretive genes is reduced.

See table \ref{tab:sims} for a summary of these results.

\begin{table}
\centering
\begin{tabular}{|c|p{6.2cm}|p{0.9cm}|p{0.9cm}|p{0.8cm}|p{0.8cm}|}
  \hline
  Scenario & Simulation description & Prob. fix. & Prob. ext. & mean $t_{fix}$ & mean $t_{ext}$ \\
  \hline
  A  & Base case & 47\%  & 53\% & 3024 & 1203 \\
  B  & Control condition, $c(S)=c(G)=0.1$. & 04\%  & 87\% & 17182 & 3062 \\
  C  & Diagonal travel & 34\%  & 66\% & 2327 & 952 \\
  D  & Epoch time $\tau=5$ & 47\%  & 53\% & 4776 & 2216 \\
  E  & Epoch time $\tau=250$ & 44\%  & 55\% & 4488 & 2007 \\
  F  & Weak Plasmid Restriction $c(S)=10^{-2}$ & 32\%  & 68\% & 4314 & 1507 \\
  G  & Strong Plasmid Restriction $c(S)=10^{-5}$ & 46\%  & 51\% & 4299 & 1778 \\
  H  & High Mutation rate ($\times 30$) & 45\%  & 55\% & 4878 & 1816 \\  
  I  & Low mutation rate $(\div 30)$ & 43\%  & 56\% & 4081 & 2470 \\
  J  & Sparse population & 11\%  & 89\% & 1595 & 440 \\
  K  & Dense population & 0\%  & 42\% & - & 11687 \\
  L  & Dense population (Death-brith process) & 46\%  & 54\% & 44433 & 17400 \\
  M  & Environment: 2 blocks & 41\%  & 58\% & 4826 & 2565 \\
  N  & Environment: Checkerboard & 43\%  & 56\% & 4836 & 1555 \\
  O  & Environment: Random & 47\%  & 51\% & 4462 & 2021 \\
  P  & Environment: Gradient & 41\%  & 59\% & 4205& 1791 \\
  Q  & Multiplasmid: linear sum \newline $f= f_a(v_a)+f_b(v_b)+f_c(v_c)$ & 45\%  & 49\% & 4600 & 2007 \\
  R  & Multiplasmid: Independent \newline $f= f_{abc}(v_a,v_b,v_c)$ & 51\%  & 46\% & 4630 & 1540 \\
  S  & Multiplasmid: Hybrid case \newline $f= f_{abc}({\bf v})+ f_a(v_a)+f_b(v_b)+f_c(v_c)$& 44\%  & 51\% & 4452 & 1810 \\
  T  & Multiplasmid: $q= 3, k=25$ & 47\%  & 52\% & 3106 & 1213 \\
  U  & Group selection. & 9\%  & 91\% & 7721 & 2438 \\
  V  & Group section, balanced initial conditions & 52\%  & 48\% & 5400 & 5850 \\
 \hline  
\end{tabular}
\caption{Table of results of simulations. For each scenario, we give a brief description, fixation and extinction probability (of secretive chromosomal genes), and the mean time to fixation or extinction. By default, initial population densities are $6\%$ secretive bacteria, $54\%$ generous (with the exception of scenario V, where we have $30\%,30\%$. Because not all simulations reach fixation before time out, probabilities do not add to $100\%$. Aside from scenario L have a time cut off of $t=20,000$.  }
\label{tab:sims}
\end{table}

\section{Conclusion}
Horizontal gene transfer amongst micro-organisms is a significant contributor to the vast complexity and variety of life we see in the world today. It allows for the sharing of resistance\cite{summers_mercury_1972,bennett_plasmid_2008} novel metabolic pathways\cite{shao_dna_2009}, and virulance factors\cite{johnson_pathogenomics_2009}, and is critical both to our understanding of life, and our forays into medicine. HGT is also something of a mystery, complicating our understanding of the evolutionary `tree'. Plasmids have been described as `paradoxical' in the litereature \cite{harrison_plasmid-mediated_2012}, with Harrison and Brockhurst observing that costly plasmids should be purged via purifying selection, while beneficial plasmid genes would be expected to integrate into the chromosome, rendering the plasmid bound copy redundant.
In this work, we present another paradox, namely the question of why the sharing of plasmids itself is not selected against.

Many past mathematical models have explored the dynamics of HGT and plasmid conjugation from the point of view of the plasmids themselves\cite{stewart_population_1977,tazzyman_fixation_2013,hall_sourcesink_2016,lili_persistence_2007,halleran_quantitative_2019,bergstrom_natural_2000,hoeven_mathematical_1984}. However, plasmids are only one half of the evolutionary story.
In this work, we have explored a number of models of HGT, focusing instead on the evolutionary dynamics of chromosonal DNA. We are able to show that for a wide array of modelling assumptions and parameter values, chromosonal genes that restrict bacterial conjugation rates have an evolutionary advantage. Even when deliberately constructing models which have classically been shown to favor cooperation, we find that selection for generous chromosomal genes is at best neutral, and in the vast majority of cases, selected against. With each change of environment or selective sweep, bacteria which horde advantageous plasmids to themselves fare better than those which share freely; this occurs despite ignoring the non-trivial costs \cite{colom_sex_2019,san_millan_fitness_2017,Ilangovan_structural_2015} of conjugation.

So where does this leave us? Horizonatal gene transfer and conjugation {\it are} ubiquitous in real world biology. As always, where modelling and observation disagree, it is the model which is lacking. There is no point shaking our fists at nature for disobeying the equations. 
Rather, our hope in this paper is to point out that bacterial generosity in the sharing of plasmids can not be taken for granted, especially when viewed from the point of view of individual chromosomal genes, which may benefit significantly from restricting plasmid spread. Regulatory mechanisms controlling bacterial conjugation serve two masters, and it is easy to produce circumstances where chromosomal and plasmid bound genes experience very different evolutionary pressures.
Unlike classical `public goods' interactions, where the good being shared is a transient physical resource, the sharing of plasmids involves the transfer of {\it information}, information which provides benefits over an evolutionary timescale. Unlike the classical question of sexual recombination, plasmid conjugation breaks any notion of symmetry between genes, partitioning the genome into mobile genes that benefit disproportionately from recombination, and immobile genes, which do not.

In terms of resolving this paradox, the work here has been exploratory rather than exhaustive, and there remain a significant number of avenues to explore. Here we study only the evolution of chromosomal genes, and ignore plasmid-chromosome co-evolution; it is entirely plausible that fast evolution of plasmids may be enough to escape any `regulatory restriction' chromosomal genes might impose. The cost of restricting conjugation itself may be high enough such that any such genes will be quickly purged from the `core genome' of a species as non-essential. When plasmids code for genes promoting public goods (such as siderophore production  \cite{lee_siderophore_2016}) then sharing of plasmids may be directly beneficial to the donating cell. What concerntration of mutualistic plasmid genes would be needed to stabilize HGT via this mechanism is as yet unknown. Alternative models of group selection may yet demonstrate the evolutionary advantage of generosity, even if the models considered here do not, or alternatively, justification may be found for reconsidering kin selection, or some other hypothesis that we have discounted as improbable in the current context.
Regardless of the exact models considered by future researchers, it is our hope to encourage discussion, and draw attention to the largely neglected role of chromosomal genes in the study of bacteria conjugation.
As demonstrated by those smallest of creatures that fill the world around us, with free sharing and recombination of information, answers to near any problem can be found.

\bibliographystyle{plain}

\begin{thebibliography}{64}
\providecommand{\natexlab}[1]{#1}
\providecommand{\url}[1]{\texttt{#1}}
\expandafter\ifx\csname urlstyle\endcsname\relax
  \providecommand{\doi}[1]{doi: #1}\else
  \providecommand{\doi}{doi: \begingroup \urlstyle{rm}\Url}\fi

\bibitem{baltrus_exploring_2013}
D.~A. Baltrus.
\newblock Exploring the costs of horizontal gene transfer.
\newblock \emph{Trends in Ecology \& Evolution}, 28\penalty0 (8):\penalty0
  489--495, Aug. 2013.
\newblock ISSN 0169-5347.
\newblock \doi{10.1016/j.tree.2013.04.002}.
\newblock URL
  \url{http://www.sciencedirect.com/science/article/pii/S0169534713001043}.

\bibitem{bell_interstellar_2015}
J.~Bell.
\newblock \emph{The {Interstellar} {Age}: {Inside} the {Forty}-{Year} {Voyager}
  {Mission}}.
\newblock Dutton, New York, New York, first edition edition, Feb. 2015.
\newblock ISBN 978-0-525-95432-3.

\bibitem{bennett_plasmid_2008}
P.~M. Bennett.
\newblock Plasmid encoded antibiotic resistance: acquisition and transfer of
  antibiotic resistance genes in bacteria.
\newblock \emph{British Journal of Pharmacology}, 153\penalty0 (Suppl
  1):\penalty0 S347--S357, Mar. 2008.
\newblock ISSN 0007-1188.
\newblock \doi{10.1038/sj.bjp.0707607}.
\newblock URL \url{https://www.ncbi.nlm.nih.gov/pmc/articles/PMC2268074/}.

\bibitem{bergstrom_natural_2000}
C.~T. Bergstrom, M.~Lipsitch, and B.~R. Levin.
\newblock Natural selection, infectious transfer and the existence conditions
  for bacterial plasmids.
\newblock \emph{Genetics}, 155\penalty0 (4):\penalty0 1505--1519, Aug. 2000.
\newblock ISSN 0016-6731.
\newblock URL \url{https://www.ncbi.nlm.nih.gov/pmc/articles/PMC1461221/}.

\bibitem{birch_kin_2015}
J.~Birch and S.~Okasha.
\newblock Kin {Selection} and {Its} {Critics}.
\newblock \emph{BioScience}, 65\penalty0 (1):\penalty0 22--32, Jan. 2015.
\newblock ISSN 0006-3568.
\newblock \doi{10.1093/biosci/biu196}.
\newblock URL
  \url{https://academic.oup.com/bioscience/article/65/1/22/2754282}.
\newblock Publisher: Oxford Academic.

\bibitem{bottery_adaptive_2017}
M.~J. Bottery, A.~J. Wood, and M.~A. Brockhurst.
\newblock Adaptive modulation of antibiotic resistance through intragenomic
  coevolution.
\newblock \emph{Nature Ecology \& Evolution}, 1\penalty0 (9):\penalty0
  1364--1369, Sept. 2017.
\newblock ISSN 2397-334X.
\newblock \doi{10.1038/s41559-017-0242-3}.
\newblock URL \url{https://www.nature.com/articles/s41559-017-0242-3}.
\newblock Number: 9 Publisher: Nature Publishing Group.

\bibitem{chen_evolution_2019}
K.~Chen, E.~W.~C. Chan, and S.~Chen.
\newblock Evolution and transmission of a conjugative plasmid encoding both
  ciprofloxacin and ceftriaxone resistance in {Salmonella}.
\newblock \emph{Emerging Microbes \& Infections}, 8\penalty0 (1):\penalty0
  396--403, Jan. 2019.
\newblock ISSN null.
\newblock \doi{10.1080/22221751.2019.1585965}.
\newblock URL \url{https://doi.org/10.1080/22221751.2019.1585965}.
\newblock Publisher: Taylor \& Francis \_eprint:
  https://doi.org/10.1080/22221751.2019.1585965.

\bibitem{colegrave_sex_2002}
N.~Colegrave.
\newblock Sex releases the speed limit on evolution.
\newblock \emph{Nature}, 420\penalty0 (6916):\penalty0 664--666, Dec. 2002.
\newblock ISSN 1476-4687.
\newblock \doi{10.1038/nature01191}.
\newblock URL \url{https://www.nature.com/articles/nature01191}.

\bibitem{colom_sex_2019}
J.~Colom, D.~Batista, A.~Baig, Y.~Tang, S.~Liu, F.~Yuan, A.~Belkhiri,
  L.~Marcelino, F.~Barbosa, M.~Rubio, R.~Atterbury, A.~Berchieri, and
  P.~Barrow.
\newblock Sex pilus specific bacteriophage to drive bacterial population
  towards antibiotic sensitivity.
\newblock \emph{Scientific Reports}, 9\penalty0 (1):\penalty0 1--11, Aug. 2019.
\newblock ISSN 2045-2322.
\newblock \doi{10.1038/s41598-019-48483-9}.
\newblock URL \url{https://www.nature.com/articles/s41598-019-48483-9}.

\bibitem{cordoni_back_2019}
G.~Cordoni and E.~Palagi.
\newblock Back to the {Future}: {A} {Glance} {Over} {Wolf} {Social} {Behavior}
  to {Understand} {Dog}–{Human} {Relationship}.
\newblock \emph{Animals : an Open Access Journal from MDPI}, 9\penalty0 (11),
  Nov. 2019.
\newblock \doi{10.3390/ani9110991}.
\newblock URL \url{https://www.ncbi.nlm.nih.gov/pmc/articles/PMC6912837/}.
\newblock Publisher: Multidisciplinary Digital Publishing Institute (MDPI).

\bibitem{cressman_replicator_2014}
R.~Cressman and Y.~Tao.
\newblock The replicator equation and other game dynamics.
\newblock \emph{Proceedings of the National Academy of Sciences}, 111\penalty0
  (Supplement 3):\penalty0 10810--10817, July 2014.
\newblock ISSN 0027-8424, 1091-6490.
\newblock \doi{10.1073/pnas.1400823111}.
\newblock URL \url{https://www.pnas.org/content/111/Supplement\_3/10810}.

\bibitem{cullum_rate_1979}
J.~Cullum and P.~Broda.
\newblock Rate of segregation due to plasmid incompatibility.
\newblock \emph{Genetics Research}, 33\penalty0 (1):\penalty0 61--79, Feb.
  1979.
\newblock ISSN 1469-5073, 0016-6723.
\newblock \doi{10.1017/S0016672300018176}.
\newblock URL
  \url{https://www.cambridge.org/core/journals/genetics-research/article/rate-of-segregation-due-to-plasmid-incompatibility/CFF64538F96DCCC0438D65AFA7A87AA3}.
\newblock Publisher: Cambridge University Press.

\bibitem{dagan_modular_2008}
T.~Dagan, Y.~Artzy-Randrup, and W.~Martin.
\newblock Modular networks and cumulative impact of lateral transfer in
  prokaryote genome evolution.
\newblock \emph{Proceedings of the National Academy of Sciences}, 105\penalty0
  (29):\penalty0 10039--10044, July 2008.
\newblock ISSN 0027-8424, 1091-6490.
\newblock \doi{10.1073/pnas.0800679105}.
\newblock URL \url{https://www.pnas.org/content/105/29/10039}.
\newblock ISBN: 9780800679101 Publisher: National Academy of Sciences Section:
  Biological Sciences.

\bibitem{dahlberg_amelioration_2003}
C.~Dahlberg and L.~Chao.
\newblock Amelioration of the cost of conjugative plasmid carriage in
  {Eschericha} coli {K12}.
\newblock \emph{Genetics}, 165\penalty0 (4):\penalty0 1641--1649, Dec. 2003.
\newblock ISSN 0016-6731.
\newblock URL \url{https://www.ncbi.nlm.nih.gov/pmc/articles/PMC1462891/}.

\bibitem{eberhard_evolution_1990}
W.~G. Eberhard.
\newblock Evolution in bacterial plasmids and levels of selection.
\newblock \emph{The Quarterly Review of Biology}, 65\penalty0 (1):\penalty0
  3--22, Mar. 1990.
\newblock ISSN 0033-5770.

\bibitem{gago-cordoba_surface_2019}
C.~Gago-Córdoba, J.~Val-Calvo, A.~Miguel-Arribas, E.~Serrano, P.~K. Singh,
  D.~Abia, L.~J. Wu, and W.~J.~J. Meijer.
\newblock Surface {Exclusion} {Revisited}: {Function} {Related} to
  {Differential} {Expression} of the {Surface} {Exclusion} {System} of
  {Bacillus} subtilis {Plasmid} {pLS20}.
\newblock \emph{Frontiers in Microbiology}, 10, 2019.
\newblock ISSN 1664-302X.
\newblock \doi{10.3389/fmicb.2019.01502}.
\newblock URL
  \url{https://www.frontiersin.org/articles/10.3389/fmicb.2019.01502/full}.
\newblock Publisher: Frontiers.

\bibitem{garcillan-barcia_why_2008}
M.~P. Garcillán-Barcia and F.~de~la Cruz.
\newblock Why is entry exclusion an essential feature of conjugative plasmids?
\newblock \emph{Plasmid}, 60\penalty0 (1):\penalty0 1--18, July 2008.
\newblock ISSN 0147-619X.
\newblock \doi{10.1016/j.plasmid.2008.03.002}.
\newblock URL
  \url{http://www.sciencedirect.com/science/article/pii/S0147619X08000231}.

\bibitem{gerrish_fate_1998}
P.~J. Gerrish and R.~E. Lenski.
\newblock The fate of competing beneficial mutations in an asexual population.
\newblock \emph{Genetica}, 102\penalty0 (0):\penalty0 127, Mar. 1998.
\newblock ISSN 1573-6857.
\newblock \doi{10.1023/A:1017067816551}.
\newblock URL \url{https://doi.org/10.1023/A:1017067816551}.

\bibitem{haldane_causes_1932}
J.~B. S. J. B.~S. Haldane.
\newblock \emph{The causes of evolution. --}.
\newblock London : Longmans, Green, 1932.
\newblock URL \url{http://archive.org/details/causesofevolutio00hald\_0}.

\bibitem{hall_sourcesink_2016}
J.~P.~J. Hall, A.~J. Wood, E.~Harrison, and M.~A. Brockhurst.
\newblock Source–sink plasmid transfer dynamics maintain gene mobility in
  soil bacterial communities.
\newblock \emph{Proceedings of the National Academy of Sciences}, page
  201600974, July 2016.
\newblock ISSN 0027-8424, 1091-6490.
\newblock \doi{10.1073/pnas.1600974113}.
\newblock URL \url{https://www.pnas.org/content/early/2016/07/05/1600974113}.

\bibitem{halleran_quantitative_2019}
A.~D. Halleran, E.~Flores-Bautista, and R.~M. Murray.
\newblock Quantitative characterization of random partitioning in the evolution
  of plasmid-encoded traits, Mar. 2019.
\newblock URL
  \url{http://resolver.caltech.edu/CaltechAUTHORS:20190402-080939441}.

\bibitem{harrison_plasmid-mediated_2012}
E.~Harrison and M.~A. Brockhurst.
\newblock Plasmid-mediated horizontal gene transfer is a coevolutionary
  process.
\newblock \emph{Trends in Microbiology}, 20\penalty0 (6):\penalty0 262--267,
  June 2012.
\newblock ISSN 1878-4380.
\newblock \doi{10.1016/j.tim.2012.04.003}.

\bibitem{hermsen_rapidity_2012}
R.~Hermsen, J.~B. Deris, and T.~Hwa.
\newblock On the rapidity of antibiotic resistance evolution facilitated by a
  concentration gradient.
\newblock \emph{Proceedings of the National Academy of Sciences}, 109\penalty0
  (27):\penalty0 10775--10780, July 2012.
\newblock ISSN 0027-8424, 1091-6490.
\newblock \doi{10.1073/pnas.1117716109}.
\newblock URL \url{https://www.pnas.org/content/109/27/10775}.
\newblock Publisher: National Academy of Sciences Section: Physical Sciences.

\bibitem{hoeven_mathematical_1984}
N.~v.~d. Hoeven.
\newblock A mathematical model for the co-existence of incompatible,
  conjugative plasmids in individual bacteria of a bacterial population.
\newblock \emph{Journal of Theoretical Biology}, 110\penalty0 (3):\penalty0
  411--423, Oct. 1984.
\newblock ISSN 0022-5193.
\newblock \doi{10.1016/S0022-5193(84)80183-6}.
\newblock URL
  \url{http://www.sciencedirect.com/science/article/pii/S0022519384801836}.

\bibitem{Hortal_role_2017}
S.~Hortal, K.~L. Plett, J.~M. Plett, T.~Cresswell, M.~Johansen, E.~Pendall, and
  I.~C. Anderson.
\newblock Role of plant–fungal nutrient trading and host control in
  determining the competitive success of ectomycorrhizal fungi.
\newblock \emph{The ISME Journal}, 11\penalty0 (12):\penalty0 2666--2676, Dec.
  2017.
\newblock ISSN 1751-7370.
\newblock \doi{10.1038/ismej.2017.116}.
\newblock URL \url{https://www.nature.com/articles/ismej2017116}.
\newblock Number: 12 Publisher: Nature Publishing Group.

\bibitem{Ilangovan_structural_2015}
A.~Ilangovan, S.~Connery, and G.~Waksman.
\newblock Structural biology of the {Gram}-negative bacterial conjugation
  systems.
\newblock \emph{Trends in Microbiology}, 23\penalty0 (5):\penalty0 301--310,
  May 2015.
\newblock ISSN 1878-4380.
\newblock \doi{10.1016/j.tim.2015.02.012}.

\bibitem{ince_ordinary_1956}
E.~L. Ince.
\newblock \emph{Ordinary {Differential} {Equations}}.
\newblock 1956.

\bibitem{institute_artificial_1989}
S.~F. Institute and C.~G. Langton.
\newblock \emph{Artificial {Life}, {Volume} {I}: {Proceedings} {Of} {An}
  {Interdisciplinary} {Workshop} {On} {Synthesis} {And} {Simulation} {Of}
  {Living} {Systems}}.
\newblock Westview Press, Redwood City, Calif, Jan. 1989.
\newblock ISBN 978-0-201-09346-9.

\bibitem{johnson_integrative_2015}
C.~M. Johnson and A.~D. Grossman.
\newblock Integrative and {Conjugative} {Elements} ({ICEs}): {What} {They} {Do}
  and {How} {They} {Work}.
\newblock \emph{Annual Review of Genetics}, 49:\penalty0 577--601, 2015.
\newblock ISSN 1545-2948.
\newblock \doi{10.1146/annurev-genet-112414-055018}.

\bibitem{johnson_pathogenomics_2009}
T.~J. Johnson and L.~K. Nolan.
\newblock Pathogenomics of the {Virulence} {Plasmids} of {Escherichia} coli.
\newblock \emph{Microbiology and Molecular Biology Reviews : MMBR}, 73\penalty0
  (4):\penalty0 750--774, Dec. 2009.
\newblock ISSN 1092-2172.
\newblock \doi{10.1128/MMBR.00015-09}.
\newblock URL \url{https://www.ncbi.nlm.nih.gov/pmc/articles/PMC2786578/}.

\bibitem{kiers_reciprocal_2011}
T.~Kiers, M.~Duhamel, Y.~Beesetty, J.~Mensah, O.~Franken, E.~Verbruggen,
  C.~Fellbaum, G.~Kowalchuk, M.~Hart, A.~Bago, T.~Palmer, S.~West,
  P.~Vandenkoornhuyse, J.~Jansa, and H.~Bücking.
\newblock Reciprocal {Rewards} {Stabilize} {Cooperation} in the {Mycorrhizal}
  {Symbiosis}.
\newblock \emph{Science (New York, N.Y.)}, 333:\penalty0 880--2, Aug. 2011.
\newblock \doi{10.1126/science.1208473}.

\bibitem{koraimann_social_2014}
G.~Koraimann and M.~A. Wagner.
\newblock Social behavior and decision making in bacterial conjugation.
\newblock \emph{Frontiers in Cellular and Infection Microbiology}, 4, Apr.
  2014.
\newblock ISSN 2235-2988.
\newblock \doi{10.3389/fcimb.2014.00054}.
\newblock URL \url{https://www.ncbi.nlm.nih.gov/pmc/articles/PMC4010749/}.

\bibitem{lee_siderophore_2016}
W.~Lee, M.~van Baalen, and V.~A.~A. Jansen.
\newblock Siderophore production and the evolution of investment in a public
  good: {An} adaptive dynamics approach to kin selection.
\newblock \emph{Journal of Theoretical Biology}, 388:\penalty0 61--71, Jan.
  2016.
\newblock ISSN 0022-5193.
\newblock \doi{10.1016/j.jtbi.2015.09.038}.
\newblock URL
  \url{http://www.sciencedirect.com/science/article/pii/S0022519315004889}.

\bibitem{lehmann_group_2007}
L.~Lehmann, L.~Keller, S.~West, and D.~Roze.
\newblock Group selection and kin selection: {Two} concepts but one process.
\newblock \emph{Proceedings of the National Academy of Sciences}, 104\penalty0
  (16):\penalty0 6736--6739, Apr. 2007.
\newblock ISSN 0027-8424, 1091-6490.
\newblock \doi{10.1073/pnas.0700662104}.
\newblock URL \url{https://www.pnas.org/content/104/16/6736}.
\newblock Publisher: National Academy of Sciences Section: Biological Sciences.

\bibitem{Lieberman_evolutionary_2005}
E.~Lieberman, C.~Hauert, and M.~A. Nowak.
\newblock Evolutionary dynamics on graphs.
\newblock \emph{Nature}, 433\penalty0 (7023):\penalty0 312--316, Jan. 2005.
\newblock ISSN 1476-4687.
\newblock \doi{10.1038/nature03204}.
\newblock URL \url{https://www.nature.com/articles/nature03204}.

\bibitem{lili_persistence_2007}
L.~N. Lili, N.~F. Britton, and E.~J. Feil.
\newblock The persistence of parasitic plasmids.
\newblock \emph{Genetics}, 177\penalty0 (1):\penalty0 399--405, Sept. 2007.
\newblock ISSN 0016-6731.
\newblock \doi{10.1534/genetics.107.077420}.

\bibitem{maynard_smith_what_1971}
J.~Maynard~Smith.
\newblock What use is sex?
\newblock \emph{Journal of Theoretical Biology}, 30\penalty0 (2):\penalty0
  319--335, Feb. 1971.
\newblock ISSN 0022-5193.
\newblock \doi{10.1016/0022-5193(71)90058-0}.
\newblock URL
  \url{http://www.sciencedirect.com/science/article/pii/0022519371900580}.

\bibitem{moran_random_1958}
P.~a.~P. Moran.
\newblock Random processes in genetics.
\newblock \emph{Mathematical Proceedings of the Cambridge Philosophical
  Society}, 54\penalty0 (1):\penalty0 60--71, Jan. 1958.
\newblock ISSN 1469-8064, 0305-0041.
\newblock \doi{10.1017/S0305004100033193}.
\newblock URL
  \url{https://www.cambridge.org/core/journals/mathematical-proceedings-of-the-cambridge-philosophical-society/article/random-processes-in-genetics/9EEED52D6AE22A026036F32D9B1CA07C}.
\newblock Publisher: Cambridge University Press.

\bibitem{norman_conjugative_2009}
A.~Norman, L.~H. Hansen, and S.~J. Sørensen.
\newblock Conjugative plasmids: vessels of the communal gene pool.
\newblock \emph{Philosophical Transactions of the Royal Society of London.
  Series B, Biological Sciences}, 364\penalty0 (1527):\penalty0 2275--2289,
  Aug. 2009.
\newblock ISSN 1471-2970.
\newblock \doi{10.1098/rstb.2009.0037}.

\bibitem{nowak_evolutionary_2006}
M.~A. Nowak.
\newblock \emph{Evolutionary {Dynamics}}.
\newblock Harvard University Press, 2006{\natexlab{a}}.
\newblock ISBN 0-674-02338-2 978-0-674-02338-3.
\newblock URL \url{http://www.hup.harvard.edu/catalog.php?isbn=9780674023383}.

\bibitem{nowak_five_2006}
M.~A. Nowak.
\newblock Five {Rules} for the {Evolution} of {Cooperation}.
\newblock \emph{Science}, 314\penalty0 (5805):\penalty0 1560--1563, Dec.
  2006{\natexlab{b}}.
\newblock ISSN 0036-8075, 1095-9203.
\newblock \doi{10.1126/science.1133755}.
\newblock URL \url{https://science.sciencemag.org/content/314/5805/1560}.
\newblock Publisher: American Association for the Advancement of Science
  Section: Review.

\bibitem{Nowak_evolution_2005}
M.~A. Nowak and K.~Sigmund.
\newblock Evolution of indirect reciprocity.
\newblock \emph{Nature}, 437\penalty0 (7063):\penalty0 1291--1298, Oct. 2005.
\newblock ISSN 1476-4687.
\newblock \doi{10.1038/nature04131}.

\bibitem{ohtsuki_how_2004}
H.~Ohtsuki and Y.~Iwasa.
\newblock How should we define goodness?--reputation dynamics in indirect
  reciprocity.
\newblock \emph{Journal of Theoretical Biology}, 231\penalty0 (1):\penalty0
  107--120, Nov. 2004.
\newblock ISSN 0022-5193.
\newblock \doi{10.1016/j.jtbi.2004.06.005}.

\bibitem{oliveira_regulation_2016}
P.~H. Oliveira, M.~Touchon, and E.~P.~C. Rocha.
\newblock Regulation of genetic flux between bacteria by
  restriction–modification systems.
\newblock \emph{Proceedings of the National Academy of Sciences of the United
  States of America}, 113\penalty0 (20):\penalty0 5658--5663, May 2016.
\newblock ISSN 0027-8424.
\newblock \doi{10.1073/pnas.1603257113}.
\newblock URL \url{https://www.ncbi.nlm.nih.gov/pmc/articles/PMC4878467/}.

\bibitem{ostrowski_enforcing_2019}
E.~A. Ostrowski.
\newblock Enforcing {Cooperation} in the {Social} {Amoebae}.
\newblock \emph{Current Biology}, 29\penalty0 (11):\penalty0 R474--R484, June
  2019.
\newblock ISSN 0960-9822.
\newblock \doi{10.1016/j.cub.2019.04.022}.
\newblock URL
  \url{http://www.sciencedirect.com/science/article/pii/S0960982219304208}.

\bibitem{otto_evolution_1997}
S.~P. Otto and N.~H. Barton.
\newblock The {Evolution} of {Recombination}: {Removing} the {Limits} to
  {Natural} {Selection}.
\newblock \emph{Genetics}, 147\penalty0 (2):\penalty0 879--906, Oct. 1997.
\newblock ISSN 0016-6731.
\newblock URL \url{https://www.ncbi.nlm.nih.gov/pmc/articles/PMC1208206/}.

\bibitem{peter_tracking_2020}
S.~Peter, M.~Bosio, C.~Gross, D.~Bezdan, J.~Gutierrez, P.~Oberhettinger,
  J.~Liese, W.~Vogel, D.~Dörfel, L.~Berger, M.~Marschal, M.~Willmann, I.~Gut,
  M.~Gut, I.~Autenrieth, and S.~Ossowski.
\newblock Tracking of {Antibiotic} {Resistance} {Transfer} and {Rapid}
  {Plasmid} {Evolution} in a {Hospital} {Setting} by {Nanopore} {Sequencing}.
\newblock \emph{mSphere}, 5\penalty0 (4), Aug. 2020.
\newblock ISSN 2379-5042.
\newblock \doi{10.1128/mSphere.00525-20}.
\newblock URL \url{https://msphere.asm.org/content/5/4/e00525-20}.
\newblock Publisher: American Society for Microbiology Journals Section:
  Research Article.

\bibitem{san_millan_fitness_2017}
A.~San~Millan and C.~Maclean.
\newblock Fitness {Costs} of {Plasmids}: a {Limit} to {Plasmid} {Transmission}.
\newblock \emph{Microbiology Spectrum}, 5, Sept. 2017.
\newblock \doi{10.1128/microbiolspec.MTBP-0016-2017}.

\bibitem{scott_regulation_1984}
J.~R. Scott.
\newblock Regulation of plasmid replication.
\newblock \emph{Microbiological Reviews}, 48\penalty0 (1):\penalty0 1--23, Mar.
  1984.
\newblock ISSN 0146-0749.
\newblock URL \url{https://www.ncbi.nlm.nih.gov/pmc/articles/PMC373000/}.

\bibitem{shao_dna_2009}
Z.~Shao, H.~Zhao, and H.~Zhao.
\newblock {DNA} assembler, an in vivo genetic method for rapid construction of
  biochemical pathways.
\newblock \emph{Nucleic Acids Research}, 37\penalty0 (2):\penalty0 e16--e16,
  Feb. 2009.
\newblock ISSN 0305-1048.
\newblock \doi{10.1093/nar/gkn991}.
\newblock URL \url{https://academic.oup.com/nar/article/37/2/e16/2410275}.
\newblock Publisher: Oxford Academic.

\bibitem{stalder_emerging_2017}
T.~Stalder, L.~M. Rogers, C.~Renfrow, H.~Yano, Z.~Smith, and E.~M. Top.
\newblock Emerging patterns of plasmid-host coevolution that stabilize
  antibiotic resistance.
\newblock \emph{Scientific Reports}, 7\penalty0 (1):\penalty0 4853, July 2017.
\newblock ISSN 2045-2322.
\newblock \doi{10.1038/s41598-017-04662-0}.
\newblock URL \url{https://www.nature.com/articles/s41598-017-04662-0}.
\newblock Number: 1 Publisher: Nature Publishing Group.

\bibitem{stewart_population_1977}
F.~M. Stewart and B.~R. Levin.
\newblock The {Population} {Biology} of {Bacterial} {Plasmids}: {A} {PRIORI}
  {Conditions} for the {Existence} of {Conjugationally} {Transmitted}
  {Factors}.
\newblock \emph{Genetics}, 87\penalty0 (2):\penalty0 209--228, Oct. 1977.
\newblock ISSN 0016-6731.

\bibitem{summers_mercury_1972}
A.~O. Summers and S.~Silver.
\newblock Mercury {Resistance} in a {Plasmid}-{Bearing} {Strain} of
  {Escherichia} coli.
\newblock \emph{Journal of Bacteriology}, 112\penalty0 (3):\penalty0
  1228--1236, Dec. 1972.
\newblock ISSN 0021-9193.
\newblock URL \url{https://www.ncbi.nlm.nih.gov/pmc/articles/PMC251553/}.

\bibitem{sun_pull_2018}
D.~Sun.
\newblock Pull in and {Push} {Out}: {Mechanisms} of {Horizontal} {Gene}
  {Transfer} in {Bacteria}.
\newblock \emph{Frontiers in Microbiology}, 9, 2018.
\newblock ISSN 1664-302X.
\newblock \doi{10.3389/fmicb.2018.02154}.
\newblock URL
  \url{https://www.frontiersin.org/articles/10.3389/fmicb.2018.02154/full}.
\newblock Publisher: Frontiers.

\bibitem{tazzyman_fixation_2013}
S.~J. Tazzyman and S.~Bonhoeffer.
\newblock Fixation probability of mobile genetic elements such as plasmids.
\newblock \emph{Theoretical Population Biology}, 90:\penalty0 49--55, Dec.
  2013.
\newblock ISSN 1096-0325.
\newblock \doi{10.1016/j.tpb.2013.09.012}.

\bibitem{thomas_plasmid_2021}
C.~M. Thomas.
\newblock Plasmid {Incompatibility}.
\newblock In E.~Bell, editor, \emph{Molecular {Life} {Sciences}: {An}
  {Encyclopedic} {Reference}}, pages 1--3. Springer, New York, NY, 2021.
\newblock ISBN 978-1-4614-6436-5.
\newblock \doi{10.1007/978-1-4614-6436-5\_565-2}.
\newblock URL \url{https://doi.org/10.1007/978-1-4614-6436-5\_565-2}.

\bibitem{thomas_mechanisms_2005}
C.~M. Thomas and K.~M. Nielsen.
\newblock Mechanisms of, and {Barriers} to, {Horizontal} {Gene} {Transfer}
  between {Bacteria}.
\newblock \emph{Nature Reviews Microbiology}, 3\penalty0 (9):\penalty0
  711--721, Sept. 2005.
\newblock ISSN 1740-1534.
\newblock \doi{10.1038/nrmicro1234}.
\newblock URL \url{https://www.nature.com/articles/nrmicro1234}.
\newblock Number: 9 Publisher: Nature Publishing Group.

\bibitem{traulsen_evolution_2006}
A.~Traulsen and M.~A. Nowak.
\newblock Evolution of cooperation by multilevel selection.
\newblock \emph{Proceedings of the National Academy of Sciences of the United
  States of America}, 103\penalty0 (29):\penalty0 10952--10955, July 2006.
\newblock ISSN 0027-8424.
\newblock \doi{10.1073/pnas.0602530103}.
\newblock URL \url{https://www.ncbi.nlm.nih.gov/pmc/articles/PMC1544155/}.

\bibitem{trautwein_native_2016}
K.~Trautwein, S.~E. Will, R.~Hulsch, U.~Maschmann, K.~Wiegmann, M.~Hensler,
  V.~Michael, H.~Ruppersberg, D.~Wünsch, C.~Feenders, M.~Neumann‐Schaal,
  S.~Kaltenhäuser, M.~Ulbrich, K.~Schmidt‐Hohagen, B.~Blasius, J.~Petersen,
  D.~Schomburg, and R.~Rabus.
\newblock Native plasmids restrict growth of {Phaeobacter} inhibens {DSM}
  17395: {Energetic} costs of plasmids assessed by quantitative physiological
  analyses.
\newblock \emph{Environmental Microbiology}, 18\penalty0 (12):\penalty0
  4817--4829, 2016.
\newblock ISSN 1462-2920.
\newblock \doi{10.1111/1462-2920.13381}.
\newblock URL
  \url{https://onlinelibrary.wiley.com/doi/abs/10.1111/1462-2920.13381}.

\bibitem{verma_revealing_2014}
T.~Verma, N.~a.~M. Araújo, and H.~J. Herrmann.
\newblock Revealing the structure of the world airline network.
\newblock \emph{Scientific Reports}, 4\penalty0 (1):\penalty0 5638, July 2014.
\newblock ISSN 2045-2322.
\newblock \doi{10.1038/srep05638}.
\newblock URL \url{https://www.nature.com/articles/srep05638}.
\newblock Number: 1 Publisher: Nature Publishing Group.

\bibitem{wade_critical_1978}
M.~J. Wade.
\newblock A {Critical} {Review} of the {Models} of {Group} {Selection}.
\newblock \emph{The Quarterly Review of Biology}, 53\penalty0 (2):\penalty0
  101--114, June 1978.
\newblock ISSN 0033-5770.
\newblock \doi{10.1086/410450}.
\newblock URL \url{https://www.journals.uchicago.edu/doi/abs/10.1086/410450}.
\newblock Publisher: The University of Chicago Press.

\bibitem{williams_sex_2020}
G.~C. Williams.
\newblock \emph{Sex and {Evolution}. ({MPB}-8), {Volume} 8}.
\newblock Princeton University Press, Mar. 2020.
\newblock ISBN 978-0-691-20992-0.
\newblock Google-Books-ID: i\_3RDwAAQBAJ.

\bibitem{williams_introduction_1971}
R.~Williams.
\newblock Introduction.
\newblock In R.~Williams, editor, \emph{Politics and {Technology}}, Studies in
  {Comparative} {Politics}, pages 7--10. Macmillan Education UK, London, 1971.
\newblock ISBN 978-1-349-01385-2.
\newblock \doi{10.1007/978-1-349-01385-2\_1}.
\newblock URL \url{https://doi.org/10.1007/978-1-349-01385-2\_1}.

\bibitem{yang_conjugative_2019}
X.~Yang, E.~Wai-Chi~Chan, R.~Zhang, and S.~Chen.
\newblock A conjugative plasmid that augments virulence in {Klebsiella}
  pneumoniae.
\newblock \emph{Nature Microbiology}, 4\penalty0 (12):\penalty0 2039--2043,
  Dec. 2019.
\newblock ISSN 2058-5276.
\newblock \doi{10.1038/s41564-019-0566-7}.
\newblock URL \url{https://www.nature.com/articles/s41564-019-0566-7}.
\newblock Number: 12 Publisher: Nature Publishing Group.

\end{thebibliography}

\end{document}